# Temporal and spatial separations between spin glass and short-range order

Margarita G. Dronova[1], Feng Ye[2], Zachary J. Morgan[2], Yishu Wang[3,4], Yejun Feng[1,5,]*

[1]Okinawa Institute of Science and Technology Graduate University, Onna, Okinawa 904-0495, Japan
[2]Neutron Scattering Division, Oak Ridge National Laboratory, Oak Ridge, Tennessee 37831, USA
[3]Department of Materials Science and Engineering, University of Tennessee, Knoxville, Tennessee 37996, USA
[4]Department of Physics and Astronomy, University of Tennessee, Knoxville, Tennessee 37996, USA
[5]Division of Physics, Mathematics, and Astronomy, California Institute of Technology, Pasadena, California 91125, USA

*Corresponding author: yejun@oist.jp

**Abstract:**

**Broken-symmetry-induced order parameters account for many phenomena in condensed matter physics. For spin glasses, such a framework dictates its theoretical construction, whereas experiments have only established dynamical behaviors such as frequency-dependent magnetic susceptibility and aging but not the thermodynamic phase. Experimental techniques have limitations when the spin glass is probed as an isolated state. To resolve this conundrum, we create an evolution from long-range order using a well-controlled tuning of the disorder on a spinel's sublattice. Cross-referencing a series of specimens at both long (milli-seconds to seconds) and short (picosecond) time scales illustrates the relationship between spin glass and long- and short-range orders. The dynamics of short- and long-range order formations are not affected by disorder, as revealed by neutron magnetic diffuse scattering, however the ranges of these orderings are changed by the introduced disorder. Across all specimens, the inflection point of the correlation length's temperature dependence fully matches with the peak in heat capacity, while spin glass can freeze either below or well above this characteristic temperature of spin order formation. Our results identify an uncorrelated coexistence of the two and attribute components of the spin glass to individual spins at domain walls between spin clusters.**

**Significance:**

**Unlike crystalline materials that are determined by symmetry, disordered states demand unconventional approaches and novel probing techniques. Spin glass is a developed topic in physics, yet its constituents are not clarified in the literature, with suggestions ranging from cluster spins to quasi-spins from vacancies. To take a refreshed approach, we comparatively examine a continuous evolution from antiferromagnetism to the disordered spin state, driven by one type of disorder with all others minimized to ensure the traceability. Using neutron magnetic diffuse scattering to record snapshots of spin correlations at pico-second time scale and other characterizations, our study reveals the identity of spin glass as individual spins that are separated from short-range orders but situated at the domain wall boundaries in between.**

**Introduction**

All long-range magnetic orders are alike, governed by symmetry groups and Landau's and Wilson's theorems of phase transitions. Spins without long-range order can be different in their own ways. Spin glasses, classical spin liquids, and quantum spin liquids are all governed by different organization principles and remain not fully understood. Theoretical treatments of spin glasses have been centralized around Edwards-Anderson replica theory and Parisi's solution of the Sherrington-Kirkpatrick model [1-4]. As a mathematically strict framework, it emphasizes a large degeneracy of ground states in a hierarchical multi-valley free-energy landscape, and a thermodynamic phase transition between paramagnetic and spin glass states [1-4]. However, it remains unclear whether and how well real systems reflect this ideal framework.

The experimental literature has not clearly defined spin glass; real systems have been heuristically categorized in various manners such as "canonical", "conventional" [1, 5-7], "glassy", "super-paramagnetic", "cluster-glass" [8-9], "anisotropic" [10-11], or "geometrically frustrated" [12-14]. Beyond frequency-dependent AC magnetic susceptibilities and aging in the time domain, it is not clear what experimental signatures are exclusively characteristic of spin glass physics. After half a century of studies, many of the early distinguishing features of spin glasses, such as amorphous vs. crystalline structure, metal vs. insulator, long- vs. short-range interaction [5], even spatial and spin dimensionalities, are now regarded as non-exclusive in determining the spin glass physics. Even disorder, as the conceptual pillar of spin glass [5], has been challenged. It was argued that "stoichiometric" spin glasses can be supported purely by frustration [12-14], when the chemical disorder is regarded insignificant at an estimated level of ~1% [12-13]. In those systems, the early emphasis of frustrated FM and AF interactions has been replaced by geometrically frustrated AF interactions on a triangular lattice [12-14]. Categorical differences aside, there remains a lack of experimental evidence identifying the microscopic contributors to the spin glass state.

Beyond low concentration spins in amorphous materials [1, 5], recent studies have explored spin glasses in the proximity of a long-range magnetic order with a high concentration of spins [11, 15-17]. As disorder suppresses the long-range order, short-range order becomes the most prominent feature of the doped systems. We note that even for dilute (<10% of total sites) spin systems, clusters of short-range spin correlation have been observed [6, 8]. The relationship between spin glass and the short-range order has been extensively explored by neutron scattering, with spin glass behavior often interpreted as dynamic fluctuations of spin clusters [6, 8, 11, 15-17]. However, the magnetic correlation temperature $T_{sro}$, determined as the *onset* temperature of the magnetic correlation length $\xi(T)$ [8, 15, 17], is about twice the peak temperature of the DC magnetic susceptibility $T_{chi}$, and has not been associated with other temperature scales. It becomes clear that experimental differentiation of spin glass physics necessarily requires a renewed approach.

Here we approach the spin glass conundrum with a refreshed, material-oriented perspective. Starting from the antiferromagnetic spinel $ZnFe_2O_4$, we create a $Zn(Fe_xGa_{1-x})_2O_4$ series with *B*-site occupancy of non-magnetic Ga ions as the leading tunable disorder. Other disorder such as Zn-non-stoichiometry and inversion disorder are collectively tuned below 1%. In this continuous series, thermodynamic behavior defined by magnetic susceptibility and heat capacity demonstrate two sensitive but surprisingly uncorrelated dependences on Ga-doping. In the disordered state, their difference represents one major intrigue of spin glass physics [1, 5, 7]. At the fast time scale of $10^{-12}$ seconds, the relationships between three forms of spin

organization— long and short-range orders, and a spin glass—are revealed by neutron magnetic diffuse scattering. Disorder does not affect the integrated spin scattering intensity but only its distribution in the reciprocal space. While the long-range order is characterized by critical fluctuations, static short-range orders along many different wavevectors nucleate over three to four spins' distances without subsequent growth. The characteristic temperature of spin clusters formation, now redefined as the *inflection* point of $\xi(T)$, matches with the peak temperature $T_C$ of the heat capacity in all systems from long- to short-range orders. In our specimens of various disorder levels, the spin glass is seen to freeze either below or above the temperature $T_C$ where spin orders develop. The components of the glass are thus identified as isolated single spins at domain walls between short-range orders, or on *A*-sites of the strongly inversion disordered spinel. Those single spins have slow dynamics and magnetic dissipation at acoustic frequencies, far separated from the cluster nucleation time scale of spin correlations. Our results resolve the long-standing discrepancy between thermodynamic characterizations and clarify the relationship between a spin glass and finite-range spin orders as an unrelated coexistence over different spatial scales with dynamics occurring over different temporal scales.

**Results**

**Controlling chemical disorder during growth**

Previously, it had been demonstrated that various types of disorder in spinel $ZnFe_2O_4$ can be fully controlled during crystal growth [18], and that the clean compound has a well-defined long-range antiferromagnetic ground state [19]. As both $ZnGa_2O_4$ and $ZnFe_2O_4$ are normal type spinels and $Ga^{3+}$ and $Fe^{3+}$ anions occupy the *B*-site with similar radii of 0.620 Å and 0.645 Å respectively under a six-fold coordination [20], the $Zn(Fe,Ga)_2O_4$ series is continuous without the disruption of any first-order structural phase transitions. Our experimental control of chemical disorder happens during the growth stage. Single crystals of $Zn(Fe,Ga)_2O_4$ were grown from a high-temperature solution of $ZnO$, $Ga_2O_3$, and $Fe_2O_3$ (Methods). This growth approach provides the freedom to adjust the relative ratios between three cations to achieve both *A*-site Zn stoichiometry and the desired Ga-doping ratio on the *B*-site (Methods). As demonstrated in Supplementary Figs. 1 and 2, non-stoichiometric disorder from either excessive or deficient Zn cations both raises $T_{CW}$ to less negative values and increases the peak value of the DC magnetic susceptibility $\chi_{DC}(T)$; only the composition with both $T_{CW}$ and $\chi_{DC}(T)$ values at the minimum leads to Zn stoichiometry on *A*-sites. From our previous study [18], Zn non-stoichiometry can be limited to below $\pm 0.2\%$ of the ideal value of unity.

Inversion disorder between *A*- and *B*-sites is directly dependent on the growth temperature $T_g$ because of entropy reasons [18]. A previous study of pure $ZnFe_2O_4$ has shown that stoichiometric single crystals with different levels of inversion disorder can be grown from $T_g$=850 °C, 1000 °C, and 1250 °C down, and the level of inversion disorder is correlated with $T_{CW}$ of -25 K, -20 K, and +10 K respectively [18]; $Fe^{3+}$- $Fe^{3+}$ exchange interaction between *A*- and *B*-sites is likely to increase $T_{CW}$ towards positive [21]. In this study, to make neutron measurements feasible, we grow $Zn(Fe_xGa_{1-x})_2O_4$ crystals from $T_g$=1000 °C with an average size ~1 mm (Supplementary Fig. 3). All of our Zn-stoichiometric $Zn(Fe_xGa_{1-x})_2O_4$ crystals have a consistent $T_{CW} \sim -20$ K (Supplementary Fig. 1), indicating that excessive disorder beyond the *B*-site doping is limited (see additional discussion below). The tuning process in our growth recipe leaves the random replacement of Fe by Ga ions on the *B*-site as the only major chemical disorder (Methods).

The crystal structure of ZnGa$_2$O$_4$ has the cubic space group $Fd\bar{3}m$, while ZnFe$_2$O$_4$ maintains the cubic $F\bar{4}3m$ space group across both the paramagnetic and antiferromagnetic phases [19]. With similar sizes between Fe and Ga ions, the lattice constant $a$ of Zn(Fe$_x$Ga$_{1-x}$)$_2$O$_4$ has a monotonic and continuous dependence on $x$ (Supplementary Fig. 1). As the lattice maintains a single phase with $F\bar{4}3m$ space group across all magnetic phases for at least $x > 0.20$, the Zn(Fe$_x$Ga$_{1-x}$)$_2$O$_4$ spinel series provides a model system to explore the spin glass evolution from the long-range order, without the complications associated with structural transitions that are present in Zn(Cr$_x$Ga$_{1-x}$)$_2$O$_4$ series [22].

**Disparate phase behaviors**

To establish the $x - T$ phase diagram, Zn(Fe$_x$Ga$_{1-x}$)$_2$O$_4$ is characterized by the macroscopic techniques of DC magnetic susceptibility $\chi_{\text{DC}}(T)$ and heat capacity $C_p(T)$ (Fig. 1). Here we focus on Fe concentration $x$ from 1.0 to 0.8, where the magnetic and thermal properties demonstrate dramatic variations and contrast. All measured $\chi_{\text{DC}}(T)$ have a divergent profile at the peak temperature $T_{\text{chi}}$, which varies from 10 to 8 K. Below $T_{\text{chi}}$, $\chi_{\text{DC}}(T)$ demonstrates differences between zero-field- (ZFC) and field-cooled (FC) conditions. $T_{\text{chi}}(x)$ drops monotonically with reducing Fe concentration $x$ but demonstrates a clear kink around $x \sim 0.97$ (Fig. 1).

In contrast to $T_{\text{chi}}(x)$ of susceptibility, the peak temperature $T_{\text{C}}(x)$ of heat capacity $C_p(T)$ has a non-monotonic dependence on $x$. $T_{\text{C}}(x)$ initially drops below $T_{\text{chi}}(x)$ by about 1.2 K at $x \sim 0.97$, then rises and eventually exceeds $T_{\text{chi}}$ by about 1.5 K at $x \sim 0.86$. The disparate behavior of $T_{\text{C}}(x)$ and $T_{\text{chi}}(x)$ is the focus point of the current study. A consistent discrepancy with $T_{\text{C}}$ higher than $T_{\text{chi}}$ by 10-30% has been regarded as a universal experimental feature of both metallic and insulating spin glasses [1]. Yet this difference has not been resolved [7] and is not compatible with the replica theory [3] which considers $T_{\text{C}}(x)$ and $T_{\text{chi}}(x)$ identical at the spin glass "transition". Instead of an individual case study of the spin glass state, here we demonstrate a continuous evolution from a long-range order to a spin glass state. The thermodynamic behavior dramatically changes for $\Delta x \sim$ 3-5% of Ga doping, indicating that the Ga doping $x$ is the effective control parameter while all other types of disorder are kept insignificant by comparison. Without the ability to reduce other disorder channels, this sensitive evolution would not be observable. Notably, without the dip structure around $x \sim 0.97$ (orange area in Fig. 1E), one can perceive $T_{\text{C}}(x)$ evolving monotonically and smoothly as a function of $x$ along the solid orange line in Fig. 1E. Instead, this non-monotonic behavior of $T_{\text{C}}(x)$ signals a change of thermodynamic nature in the spins' organizational principles.

**Dissipation signatures of disorder**

While AC magnetic susceptibility is a common technique to verify spin glass states, the Zn(Fe$_x$Ga$_{1-x}$)$_2$O$_4$ series provides an additional freedom to continuously track effects of disorder. Here we focus on the imaginary component of AC magnetic susceptibility $\chi''(T)$; the real part $\chi'(T)$ is of limited utility for distinguishing between antiferromagnetism and spin glasses, as their shapes are similar [1, 23]. The onset (instead of the peak) of non-vanishing $\chi''(T)$ indicates the existence of an out-of-phase response to the driving field, which bears similar physics to the ZFC vs. FC separation below $T_{\text{chi}}$ in the DC magnetic susceptibility.

Unlike in ferromagnets and metallic systems, $\chi''(T)$ typically bears no feature in insulating antiferromagnets and paramagnets but can have non-vanishing spectral weight due to disorder [23]. The peak seen in $\chi''(T)$ near $T_{\text{chi}}$ (Fig. 2) thus can be taken as a measure of the amount spin disorder in ZnFe$_2$O$_4$ under different preparation conditions. Previous characterizations [18] have set the non-stoichiometry within $\sim \pm 0.2\%$, and the inversion disorder in 1000 °C grown crystals at about two to three times higher than that of 850 °C grown crystals. Here the $\chi''(T)$ peak near $T_{\text{chi}}$ provides an assessment of the total disorder. The peak in $\chi''(T)$ at $T_{\text{chi}} \sim 10$ K is only $\sim 0.12\%$ of $\chi'(T_{\text{chi}})$ for 850 °C grown crystals (Supplementary Fig. 4C) but increases by about three times to $\sim 0.4\%$ of $\chi'$ for 1000 °C grown crystals (Supplementary Fig. 4B). The value of $\chi''/\chi'$ at around $T_{\text{chi}}$ can reach $\sim 2.2\%$ in our 1250 °C grown crystals (Supplementary Fig. 4A), and $\sim 3.3\%$ for 1200 °C grown crystals in Ref. [24]. The total disorder of non-stoichiometry and inversion types in 850 and 1000 °C grown crystals can thus be estimated as $\sim \pm 0.25\%$ and $\sim \pm 0.8\%$ of all $B$-sites respectively. These disorder levels in two stoichiometric ZnFe$_2$O$_4$ crystals justify the clean-limit and the starting points of our Ga-doping series. Assuming $A$-site Zn-stoichiometry and the level of inversion disorder remain the same, Ga doping on $B$-sites, at several percent level, is the only significant source of disorder.

In stoichiometric ZnFe$_2$O$_4$ crystals, there exists a broad feature in $\chi''(T)$ centered at $\sim 4$ K (Figs. 2A, 2B). As there exists no additional thermodynamic phase below 10 K [19], this feature might be associated with the antiferromagnetism in ZnFe$_2$O$_4$ but not a reentrant spin glass, whereas the peak feature at $T_{\text{chi}}$ represents disordered spins. For Zn(Fe$_x$Ga$_{1-x}$)$_2$O$_4$ of $x = 0.97$ and 0.95 (Figs. 2C, 2D), both spectral features in $\chi''(T)$ remain, but the peak grows more prominent. As the 4 K hump disappears with Ga-doping below $x \sim 0.9$ (Fig. 2E), $\chi''(T)$ becomes a step function form that is typical of spin glasses [1, 25]. Taking $\chi''(T)$ of $x = 1.0$ (Figs. 2A) and $x = 0.69$ (Fig. 2H) as the intrinsic behaviors of antiferromagnetism and spin glass respectively, all of the $\chi''(T)$ curves can be considered as a mixture of these two limits with different proportions. This continuous mixing indicates that the Zn(Fe$_x$Ga$_{1-x}$)$_2$O$_4$ series does not have a clear separation point in $x$ between the long-range order and the spin glass state. Even for stoichiometric crystals grown at 850 and 1000 °C, signatures of spin glass exist despite the dominant long-range order.

**Antiferromagnetic fluctuations vs. spin correlation**

To parse the role of disorder, we turn to quasi-elastic neutron scattering to provide access to faster spin dynamics and microscopic length scales. Our measured quasi-elastic diffuse scattering at CORELLI represents elastic scattering of either thermally excited spin fluctuations or static spin correlations (Methods), while the much weaker inelastic scattering signals are buried under the correlated noise of the elastic signal because of the usage of a correlation chopper [26, 27]. The quasi-elastic CORELLI setup has a relatively broad bandwidth; the detected fluctuations are thus those with time constants longer than $10^{-12}$ seconds (Methods). Quasi-elastic neutron scattering originally could only probe limited reciprocal space positions [28]. Modern diffuse scattering setups now cover the whole reciprocal space, allowing a comprehensive understanding of all magnetic correlations [6, 18]. Our neutron magnetic diffuse scattering explored three sets of single crystal arrays of $x=1.00$, 0.97, and 0.86 (Supplementary Figs. 5-6) which correspond to key compositions in the phase diagram (Fig. 1E). All spectra are inspected in Fig. 3 and Supplementary Fig. 7 through the incremental change of magnetic diffuse scattering $I_{T1-T2}(\boldsymbol{q}) = I_{T1}(\boldsymbol{q}) - I_{T2}(\boldsymbol{q})$ between two adjacent temperatures $T1$ and $T2$. The magnetic diffuse scattering is integrated over several different reciprocal space volumes and plotted in Fig. 4.

For all three compositions, we start by examining the total $q$-integrated spectral intensities normalized by the total $Fe^{3+}$ amount (Methods, Fig. 4A), which represent spin correlations that are stable within our measurement time window (~$10^{-12}$ s). These intensities are similar to each other at all temperatures, and their shapes and magnitudes show no sensitivity to the levels of disorder, the thermodynamic phase of order vs. disorder, and their differing temperature scales of $T_{\text{chi}}$. The increase of magnitude by ~3.5× from 60 to 4.8 K can be attributed to spin correlations that are less destroyed by reduced thermal energy at lower temperatures. Consequently, we can infer that disorder does not affect spin dynamics at the time scale of $10^{-12}$ seconds. However, the spectral weights grow and are distributed differently in reciprocal space depending on their respective disorder conditions.

For $ZnFe_2O_4$, the increase of spin diffuse scattering from 60 K to 10 K can be viewed as a characteristic of softened fluctuations and critical slowdown approaching a second order magnetic phase transition. For $Zn(Fe_xGa_{1-x})_2O_4$ with $x$=0.97 and 0.86 (Figs. 3B, 3C), their increased diffuse spectral weights cannot be distinguished from that of $ZnFe_2O_4$; indeed, diffuse spectra of all three systems are identical at a common temperature of 15 K (Supplementary Figs. 5, 6). Quantitatively, for integration volumes of various sizes and positions in the reciprocal space (Fig. 4), $I_q(T)$ from 60 K to 10 K manifests no difference between three samples with very different disorder levels. Again, while spins experience the correlation effect in the paramagnetic phase, as demonstrated by the Curie-Weiss law, disorder does not directly affect their dynamics at a time scale of $10^{-12}$ seconds.

Below 10 K, however, the spectra differ. For $ZnFe_2O_4$, $I_q(T)$ at different $q$ positions have opposite evolutions in temperature (Figs. 3, 4B, 4D); the spectral weight along the Brillouin zone ridge between (0.5, 1, 0) and (1, 0.5, 0) is shifted to the magnetic order vector (1, ½, 0). The integrated intensity $I_{(0.75,0.75,0)}(T)$ around reciprocal space position (¾, ¾, 0) peaks at $T_{\text{chi}}$ (Fig. 4D) and is similar in shape to $\chi_{\text{DC}}(T)$ and $\chi'(T)$ (Fig. 1A, Supplementary Fig. 4B), which is consistent with our understanding of critical fluctuations and the formation of a long-range order. For $Zn(Fe_{0.97}Ga_{0.03})_2O_4$, the temperature evolution of $I_q(T)$ is similar. Below ~9 K, $I_q(T)$ demonstrates the presence of critical fluctuations as the spectral weight at (¾, ¾, 0) is reshuffled (Fig. 4D) to the magnetic order vector (1, ½, 0), which increases significantly over a small reciprocal space volume (Fig. 4C). In $Zn(Fe_{0.86}Ga_{0.14})_2O_4$, $I_q(T)$ continues to grow with reducing temperatures for all $q$ modes, and the temperature evolutions at different $q$ are all similar. The lack of spectral weight reshuffling during the temperature evolution, for all time scales longer than $10^{-12}$ seconds, is one of the major differences between long- and short-range orders.

All integrated $I_q(T)$ of both $Zn(Fe_{0.97}Ga_{0.03})_2O_4$ and $Zn(Fe_{0.86}Ga_{0.14})_2O_4$ reveal that $I_q(T)$ has no correlation to $T_{\text{chi}}$ over various reciprocal space positions and volumes (Fig. 4). Previously, a neutron magnetic diffuse scattering of an amorphous metallic spin glass also reported insensitivity of $I_q(T)$ to $T_{\text{chi}}$ at one transferred momentum $q = 0.13$ Å$^{-1}$, with a $10^{-11}$ or $10^{-9}$ s temporal threshold for the spectral integration [28]. Here, the complete series of crystals provides a significant advantage for understanding the role of disorder, further assisted by measurements over the whole reciprocal space for various integration volumes.

**Thermal evolutions of the spin correlation length**

To search for physical signatures that can be related to either $T_{chi}$ or $T_C$, we extract temperature evolutions of spin correlation length $\xi$ of all samples (Methods). In all three systems, $\xi_{(1,0.5,0)}(T)$ are similar from 25 to 10 K (Fig. 5A). From 10 K down, $\xi_{(1,0.5,0)}$ of the long-range antiferromagnetic order in $ZnFe_2O_4$ quickly increases to a length scale ~30 Å that is identical to the correlation lengths of the lattice (Fig. 5A). For $Zn(Fe_{0.86}Ga_{0.14})_2O_4$, $\xi_{(1,0.5,0)}$ grows slowly from 5 to 11 Å as temperature drops from 15 to 4.8 K. For $Zn(Fe_{0.97}Ga_{0.03})_2O_4$, $\xi_{(1,0.5,0)}$ at 8.9 K is similar to that of $Zn(Fe_{0.86}Ga_{0.14})_2O_4$. From about 7.5 K, $\xi_{(1,0.5,0)}$ starts to grow towards the eventual correlation length of ~23 Å. We note that it does not reach the instrument resolution limit at 4.8 K, which is ~40 Å judging by the sample lattice (Methods).

For all three specimens, a smooth curve through data points captures the essence in the temperature evolution of $\xi_{(1,0.5,0)}(T)$ (Fig. 5). In Fig. 5A, each curve is also compared to peak temperature position of the heat capacity. We see clearly that for all three systems, $T_C$ is aligned with the inflection point of $\xi_{(1,0.5,0)}(T)$. The spin correlation length $\xi_{(1,0.5,0)}$ thus clarifies the meaning of the heat capacity peak temperature; the development of spin correlations has a thermodynamic implication that is related to the entropic nature of the heat capacity. Similarly, diffuse scattering intensities $I_{(1,0.5,0)}(T)$ (Fig. 4C) also show a correlation between their inflection points of the thermal evolution and each sample's respective $T_C$. More precisely, this correlation can be between two matching inflection points of entropy $S(T)$ and $\xi(T)$. The inflection point of entropy $S(T)$ and the peak of $C_P$ are expected to be identical for an infinitely sharp $C_P$ (Fig. 1A) but can have a fine difference when $C_P$ is broad. Our experimentally measured $\xi(T)$ are too coarse to distinguish between these two scenarios.

Magnetic diffuse scattering also exists prominently between a pair of (1, 0.5, 0) and (0.5, 1, 0) points, along the ridges of the first Brillouin zone of the cubic unit cell (Fig. 3). The concentrated spectral weight at the Brillouin zone boundary, such as (0.75, 0.75, 0), indicates that they are short-range correlations. The extracted spin correlation length $\xi_{(0.75,0.75,0)}(T)$ from longitudinal cuts is plotted in Fig. 5B. For $ZnFe_2O_4$ and $Zn(Fe_{0.97}Ga_{0.03})_2O_4$, both correlation lengths $\xi_{(0.75,0.75,0)}$ peak at ~9 Å, at $T_C$, before reduce to shorter lengths at lower temperatures along with drops in the spectral intensity (Fig. 4d). This suggests that the spectral weights around (0.75, 0.75, 0) are from fluctuations, and they do not turn into a magnetic instability like that at (1, 0.5, 0). For $Zn(Fe_{0.86}Ga_{0.14})_2O_4$, $\xi_{(0.75,0.75,0)}$ grows continuously to 10 to 11 Å at low temperature, identical in behavior to $\xi_{(1,0.5,0)}$ at all temperatures (Fig. 5). Our results suggest a disordered spin system can have the inhomogeneity that many similarly sized short-range orders along different wave vectors exist in respective local regions. On the *B*-site pyrochlore lattice, edges of Fe tetrahedra are all along the (1, 1, 0) family of directions of the cubic spinel lattice (Fig. 1E inset). Given that the nearest neighbor Fe-Fe distance is ~3.0 Å [19], a spin correlation length $\xi_{(0.75,0.75,0)}$ of 6 to 11 Å thus represents a group of three to four correlated $Fe^{3+}$ ions along the (1, 1, 0) direction.

**Discussion**

Previous neutron elastic scattering measurements of spin glasses have indicated that the short-range order formation temperature $T_{sro}$, often arbitrarily marked at the *onset* temperature of magnetic scattering intensity $I(T)$ [8, 15-17], leads to $T_{sro} \gg T_{chi}$ and no correlation to other thermodynamic quantities. While the discrepancy between heat capacity peak $T_C$ and susceptibility peak $T_{chi}$ is one long-standing issue in the spin glass literature [5, 7], here our results connect the origin of heat capacity to the forming of short-range order by identifying $T_\text{C}$, defined at the peak of $C_p(T)$, with $T_\text{sro}$ defined at the *inflection* point of $\xi(T)$. What is unexpected is that, even in Zn(Fe$_{0.97}$Ga$_{0.03}$)$_2$O$_4$ where $T_\text{C}$ is significantly lower than $T_\text{chi}$, $T_\text{sro}$ still closely tracks $T_\text{C}$. As heat capacity measures the entropy associated with spin configurations, $C_p$ is necessarily sensitive to the spatial range $\xi$ of spin order, which forms over time scales much shorter than the picosecond neutron observation time so that our measurement could reveal their correlation.

In the La$_{2-x}$Sr$_x$CuO$_4$ series [15], $T_\text{sro}(x)$ follows a monotonic $x$-dependence that is almost $\sim 2T_\text{chi}(x)$, and hence the spin glass state was considered to be equivalent to the short-range order. Here, through a well-controlled pathway from order to disorder in the Zn(Fe$_x$Ga$_{1-x}$)$_2$O$_4$ series, the oscillating $T_\text{C}(x)$ (equivalently $T_\text{sro}$) around $T_\text{chi}(x)$ (Fig. 1) indicates that the spin glass behavior, defined by AC frequency-dependent susceptibilities, is uncorrelated to short-range spin orders. The short correlation length of 6 to 11 Å in Zn(Fe$_{0.86}$Ga$_{0.14}$)$_2$O$_4$, about three or four spins, suggests uncorrelated individual spins as the most plausible component of the spin glass. The spin glass can be considered as an ensemble of single spins at positions such as the domain wall between short-range spin clusters. This uncorrelated relationship between single spins and spin clusters is especially clear in Zn(Fe$_{0.97}$Ga$_{0.03}$)$_2$O$_4$. At $T_\text{chi} \sim 8.9$ K which is higher than $T_\text{C} \sim 7.7$ K, the spin correlation length $\xi(T_\text{chi}) \sim 9.0$ Å is much shorter than the saturated correlation length $\xi(4.8K) \sim 23$ Å. So, single spins belonging to the spin glass freeze at a higher temperature before the long-range order is fully developed. Conversely, those spins that form the long-range order do not freeze at $T_\text{chi}$ but instead configure into the cluster arrangement only at lower temperature. A finite correlation length in Zn(Fe$_{0.97}$Ga$_{0.03}$)$_2$O$_4$ indicates ample domain walls and in turn, many single spins. In general, single spins jammed between spin clusters can experience very different energy landscapes in comparison to spins inside a cluster which explains the large time constants of their dynamics and the broad distribution.

Even in stoichiometric ZnFe$_2$O$_4$, there exist various levels of glassiness associated with different amounts of inversion disorder (Figs. 2A-2B, Supplementary Fig. 4). In addition to domain walls, a high level of inversion disorder in spinel materials can also explain the strong spin glass tendency. In ZnFe$_2$O$_4$ crystals grown at high temperatures with strong spin glass signatures (Supplementary Fig. 4A) [24], the level of disorder on the *B*-site can reach $\sim$5% and the inversion disorder allows Fe$^{3+}$ spins to reside on the *A*-site as quasi-free single spins despite the crystals being stoichiometric [18]. Both scenarios create loose spins and place a spin glass as a separated coexistence to either long- or short-range spin order.

Multiple spin correlations of different wave vectors have been observed in many spin glass systems [6, 9, 15-16]. For Mn-doped Cu and Pt single crystals at a dilute level of 5 to 16 at. %, there are signs of multiple antiferromagnetic short-range orders in addition to the ferromagnetic order [6, 9]. In spin glasses of single crystal (La,Sr)$_2$CuO$_4$, there exist both commensurate and incommensurate spin correlations [15-16]. Here, we have demonstrated that the non-monotonic evolution of $T_\text{C}(x)$ can be explained by the transitioning from one to multiple spin correlation instabilities.

In ZnFe$_2$O$_4$ and Zn(Fe$_{0.97}$Ga$_{0.03}$)$_2$O$_4$, there exists only one spin-correlation instability at the (1, 0.5, 0) wave vector, with spectral weight shifted from other reciprocal space positions to form the long-range order. A small amount of disorder elongates the buildup process of spin correlation over the reducing temperature, as some disorder-weakened links in spin correlation can only be stabilized against reduced thermal energy. $T_\text{C}$ of the $x = 0.97$ sample thus is at a lower temperature than that of clean ZnFe$_2$O$_4$. In Zn(Fe$_{0.86}$Ga$_{0.14}$)$_2$O$_4$, there are multiple spin correlation instabilities at many different wave vectors which can be distributed almost continuously from (1, 0.5, 0) to (0.75, 0.75, 0) (Fig. 3C). All the spin correlation lengths $\xi_q(T)$ evolve similarly to $\xi_{(1,0.5,0)}(T)$ with identical $T_{\text{sro}}$ (Fig. 5). These multiple instabilities of different wavevectors means that no single $\xi_q$ grows to a large size below 10 K. Instead, there exist many nucleation centers, and all clusters easily saturate in size with reducing temperature because of the high level of disorder. The short-range ordered state can hence be regarded as the result of a jammed nucleation process that prevents subsequent growth to a long-range order. As the magnetic entropy is quickly locked down, the inflection point of $\xi_q(T)$ will occur at a high temperature. More precisely, the characteristic temperature of the short-range order should be identical to that of the disorder-free long-range order as both involve the nucleation process that mainly depends on local exchange strength. Only in the $x = 0.97$ sample, the gradual development of the long-range order over weakened links demands a lower temperature, thus creating the V-shaped dip in the $T$-$x$ phase diagram. The presence of multiple instabilities can be regarded as a key characteristic of the short-range order versus the single instability of the long-range order.

Major magnetic signatures of spin glasses such as frequency-dependent AC susceptibility, FC/ZFC DC susceptibilities, and aging, can all be attributed to slow single spin relaxation that is expressed as delayed responses to external disturbances from the MHz to the sub-Hz range. The relaxation time can extend over several decades because of subtle variations in the local energy barriers, similar to single anion inversion tunneling in molecules such as $NH_3$, $ND_3$, and $PH_3$ [29]. The non-vanishing dissipation of $\chi''(T)$ supports such a picture of classical relaxation instead of quantum coherent tunneling. Our understanding of the spin glass from the dynamic perspective is similar to that of Ref. [30]. However, instead of pointing to single spins, Ref. [30] does not separate spin glass from the short-range order. Our experimental results instead demonstrate that short-range order and single spins of spin glass are uncorrelated topics over different length and time scales.

**Methods:**
**Single crystal growth.** Our single crystals were grown using the high temperature solution technique previously developed in Ref. [18]. For dissolved ZnO, $Ga_2O_3$, and $Fe_2O_3$ powders in the high-temperature solvent of anhydrous borax, there are two adjustable parameters, namely, the molar ratio $r$ of ZnO vs. the total amount of $Fe_2O_3$ and $Ga_2O_3$, and $Ga_2O_3$'s concentration $y$ (from 0 to 1) in the combined $Fe_2O_3$ and $Ga_2O_3$ powder. They together provide the tunability of the whole crystal series. Our Ga-doped $Zn(Fe,Ga)_2O_4$ crystals in this study were all grown from $T_g$=1000 °C to 850 °C, while pure $ZnFe_2O_4$ crystals were all grown previously in the study of Ref. [18].

For each growth under the condition $(r, y)$, the Curie-Weiss temperature $T_{\text{CW}}$ was extracted from measured $M(T)$ of individual crystals (Supplementary Fig. 1A) and statistics were built over multiple (~10) specimens. Supplementary Fig. 1B demonstrates $T_{\text{CW}}$ vs. $r$ for four different $y$ conditions. Less ZnO in the solution causes excessive Fe on the $A$-site and dramatically increases the Curie-Weiss temperature $T_{\text{CW}}$ toward positive values; excessive ZnO in the solution leads to extra Zn on the $B$-site and a slowly increasing $T_{\text{CW}}$. For each fixed $y$, the composition $r_{stoich}$ with the lowest $T_{\text{CW}} \sim$ -20K represents the condition for growing Zn-stoichiometric crystals. The Fe concentrations $x$ in the $Zn(Fe_xGa_{1-x})_2O_4$ crystals were determined by the slopes of the Curie-Weiss plots.

Similarly, non-stoichiometry disorder from either excessive or deficient Zn cations enhances the magnitude of the DC magnetic susceptibility $\chi_{\text{DC}}(T)$; the evolution of $\chi_{\text{DC}}(T)$ at six $r$ compositions of $y = 0.5$ are shown in Supplementary Fig. 2, with $r_{stoich}$=0.8 represents the condition for the lowest $\chi_{\text{DC}}(T)$. To grow Zn-stoichiometric crystals at other Ga concentrations, interpolated values from the smooth curve $r_{stoich}(y)$ were used (Supplementary Fig. 1C).

**DC Magnetic susceptibility.** The DC magnetic susceptibility $\chi_{\text{DC}}(T)$ was measured in a magnetic property measurement system (MPMS3, Quantum Design Inc.). Individual single crystals were cooled to 1.8 K either in zero field (ZFC) or under a magnetic field of 100 Oe (FC) and were measured during the warming process under a 100-Oe field. Before the measurement, the remanent field in the MPMS magnet was cleansed to about 1 Oe by field oscillation. $T_{\text{CW}}$ was calculated by a linear fit to the Curie-Weiss law over the high-temperature portion of the inverse susceptibility $1/\chi$. The temperature range of the fitting was determined by gradually expanding the lower boundary of the fitting range from 400 K down, to the extent that the quality of the linear fit could be either improved or maintained. The temperature range of fitting is always above 5| $T_{\text{CW}}$ |, which justifies the Curie-Weiss analysis. For each synthesis, 8 to 10 crystals were individually measured for the statistical average of $T_{\text{CW}}$.

**Heat capacity.** For each batch of growth, heat capacity $C_p(T)$ of at least two pieces of single crystals were individually measured at zero field, using a physical property measurement system (Dynacool-9, Quantum Design Inc.). The measurements used the pulse-relaxation scheme, with a temperature rise of 2% of each temperature. At each temperature, $C_p$ was measured at least three times for temperature stability, consistency, and statistical averaging.

**AC magnetic susceptibility.** AC magnetic susceptibility was measured using the measurement option of a physical property measurement system (ACMS II, Dynacool-12, Quantum Design Inc.). The ACMS II module was operated in the three-point measurement mode. For each composition of Zn(Fe$_x$Ga$_{1-x}$)$_2$O$_4$, we mounted several pieces of crystals with a total mass of 15-35 mg on a half-cylinder quartz holder to increase the measured $\chi''(T)$ signal. Phase sensitive $\chi'(T)$ and $\chi''(T)$ were measured at three frequencies of 100, 1001, and 9984 Hz, with a 15, 15, and 3 Oe driving field and a time constant of 8, 8, 6 seconds respectively. At each temperature point, 20-120 measurements were accumulated to build the statistics of $\chi''(T)$. We measured $\chi'(T)$ and $\chi''(T)$ data from 1.8 K to 17 K, with the temperature range mainly limited by an instrument induced anomaly in $\chi''(T)$ at 20-30 K, originating from an Inconel component in the ACMS coil holder. The phase zero was calibrated by a powder pellet of Yb$_2$Sn$_2$O$_7$, which has an antiferromagnetic order transition at 165 mK and zero $\chi''(T)$ over our measurement temperature range.

**Neutron magnetic diffuse scattering.** We performed neutron magnetic diffuse scattering at CORELLI, BL-9 of the Spallation Neutron Source at the Oak Ridge National Laboratory in the USA. At CORELLI, a white band of pulsed neutron flux and a statistical cross-correlation chopper are utilized to perform elastic scattering using a major part ($= 127/255 \approx 50\%$) of the incident white beam [26, 27] for enhanced counting statistics.

In general, the measured neutron scattering signals are of quasi-elastic nature and can be due to either elastic scattering capturing thermally excited spin fluctuation modes or inelastic scattering of low-energy excitations within the bandwidth $\Delta E$ of elastic energy resolution. In the elastic case, if a thermally excited fluctuation mode remains static over the time $\tau$ that neutrons fly across the distance of their coherent length $\lambda^2/\Delta\lambda$, then the thermally excited mode would generate a neutron diffraction event at the mode's wave vector $q$; faster fluctuations would be viewed as averaged out by neutrons. In the inelastic case, only inelastic scatterings with both the energy loss and inverse lifetime narrower than $\Delta E$ of the energy resolution would significantly contribute to the measured elastic line intensity. These two different origins of quasi-elastic diffuse scattering have an identical cutoff time $\tau$, which is related to the elastic bandwidth $\Delta E$ by the uncertainty relationship $\tau \cdot \Delta E \sim \hbar$.

A feature of the cross-correlation chopper scheme is that both the energy resolution $\Delta E$ and the ratio $\Delta E/E$ increase with increasing incident neutron energy $E$ [27]. CORELLI utilizes neutrons from 10 to 130 meV, with peak flux at 45 meV. The instrument has $(\Delta E_{\text{FWMH}})/E \sim$ 2%, 4%, and 6.5% at $E\sim$10, 45, 130 meV respectively, corresponding to $\Delta E_{\text{FWMH}} \sim 0.2$, 1.8, and 8.5 meV. With the usage of a statistical chopper, the measured signal at the detector can be processed through the statistical correlation procedure to remove inelastic signals outside of this $\Delta E_{\text{FWMH}}$ range. Nevertheless, the cross-correlation processed signal still contain both the elastic and inelastic signals within the energy resolution. Despite the bandwidths for high energy neutrons being relatively broad, the inelastic signals remain much weaker in comparison to the elastic signals. Furthermore, the correlation-based specific averaging scheme introduces strong statistical noises of the high intensity features (elastic scattering) across different channels, so weak inelastic signals are masked by the statistical noises of elastic signals [26]. Treating this as a technical advantage, the cross-correlation chopper scheme, such as the setup at CORELLI, is ideal for the exclusive measurement of diffuse scattering [26]. For $ZnFe_2O_4$, previous inelastic neutron scattering [31] has demonstrated that spin excitations in the pure system reside below 4 meV and peak around 1 meV. Given that our measurement temperature ranges mostly from 4.8 K to 15 K (0.5-1.5 meV), a significant fraction of the spin fluctuation modes can be populated by thermal excitations, and those thermally populated modes are accessible to our elastic scattering if their lifetimes are beyond our measurement time scale. Based on the energy resolution of the spectrometer, the time scale that the neutron diffuse scattering can be calculated in a manner similar to the procedure in Ref. [28]. We estimate the time scale $\tau = \hbar/\Delta E_{\text{HWHM}}$, and take $\Delta E_{\text{HWHM}} = 0.9$ meV for the 45 meV energy neutrons at the peak of the incident flux, so $\tau = 7 \times 10^{-13} \sim 10^{-12}$ seconds.

We assembled $Zn(Fe_xGa_{1-x})_2O_4$ crystals in a mosaic array form on both sides of thin (0.3 mm) aluminum plates and aligned several plates to form a sample assembly for three concentrations $x$=1.00, 0.97 and 0.86 (Supplementary Fig. 3). For $x$=0.97 and 0.86, single crystals of total masses 322 and 321 mg respectively were used, each collected from a single growth batch; each mosaic assembly was built within a volume of 8×8×8 mm$^3$. For $x$=1.00, crystals from three separate growth batches were used to accumulate a total mass of 1.095 gram. The assembly also had a larger volume of 10×10×10 mm$^3$. All result the lattice reflection (1, 1, 1) linewidth of the $x$=1.00 assembly slightly broader than those of assemblies of $x$=0.97 and 0.86 (Fig. 5).

The detected scattering signals were first converted into intensities as a function of ($H$, $K$, $L$) positions in the three-dimensional reciprocal space, using standard deduction software developed at the CORELLI beamline. The statistical cross-correlation procedure was applied at this stage so all data we discuss in the current work (Figs. 3, 4, 5 and Supplementary Figs. 5, 6, 7) are of elastic type [Ye2018]. Subsequent data analysis was carried out using the MANTID software interfaced with customized Python scripts. Two-dimensional plots of diffuse scattering data in the ($H$, $K$, $L$=const.) plane (Supplementary Figs. 5, 6) have the intensities integrated over a thickness of 0.10 r.l.u along the $L$ direction and normalized by moles of Fe ions of each sample. They further build the differential evolution plots of $I_{T1-T2}(\boldsymbol{q}) = I_{T1}(\boldsymbol{q}) - I_{T2}(\boldsymbol{q})$ between neighboring temperatures $T1$ and $T2$ in Fig. 3 and Supplementary Fig. 7.

For studies of integrated intensities, intensity data over the reciprocal space was binned with a resolution of 0.01 r.l.u. along all three dimensions of *H*, *K*, and *L*. We integrated over spherical volumes around specific reciprocal space points such as (1, 0, ½), (1.5, 1, 0), and (0.75, 0.75, 0) with various radii (Fig. 4 schematics). An integration includes all permutated indices for more precise averaging and to account for three different magnetic domains. For the total magnetic scattering intensities plotted in Fig. 4A, we used an integration radius of 0.354 r.l.u., which allows a no-gap and no-overlap coverage of the space between two magnetic reflection points (Fig. 4 schematics). We refrained from an integration over the whole reciprocal space, due to issues of varying and unaccountable background (such as at small *q*) and incomplete coverage over certain portions of reciprocal space (*e.g.* Fig. 3 of Ref. [18]). The background was treated as a constant and estimated by averaging over seven inequivalent positions within the first two Brillouin zones, and over the uniform space near the lattice Bragg reflections where the magnetic intensity is absent; at each position, a spherical volume integration with a radius of 0.050 r.l.u. was performed. The integration of magnetic intensity with background subtracted was normalized by the total moles of Fe ions for each sample.

For studies of the correlation length, radial line cuts for each crystal assembly and each temperature were generated for lattice (1, 1, 1) reflection, magnetic (0.5, 1, 0) reflection, and interstitial (0.75, 0.75, 0) positions. For lattice (1, 1, 1) reflection, a rectangle cross-section of $0.42 \times 0.36$ Å$^{-2}$ was used for the integration, spanning the transverse (1, -1, 0) and (1, 1, -2) directions respectively. For magnetic (0.5, 1, 0) reflection, a cross section of $0.33 \times 0.30$ Å$^{-2}$ was used for the integration, spanning the transverse (-2, 1, 0) and (0, 0, 1) directions respectively. For the interstitial (0.75, 0.75, 0) position, a cross-section of $0.21 \times 0.15$ Å$^{-2}$ was used for the integration, spanned along the transverse (-1, 1, 0) and (0, 0, 1) directions respectively. All three reciprocal space positions were chosen for their similar values of transferred momentum, leading to similar instrument resolutions. We note the instrument resolution limit of correlation length at ~40 Å is generated from the cross-correlation-deducted, three-dimensional reciprocal space intensity data, which includes the conversion from all wavelengths and scan frames. While this sets a relatively limited upper boundary, it does not affect values of all shorter correlation lengths that are deduced from the three-dimensional intensity data.

**Figure captions:**
**Fig. 1. Disparate phase behaviors of Zn(Fe,Ga)$_2$O$_4$.** (A-D) DC magnetic susceptibility $\chi_{\text{DC}}(T)$ and heat capacity $C_\text{p}(T)$ of Zn(Fe$_x$Ga$_{1-x}$)$_2$O$_4$ are plotted in the same panel to illustrate the evolving difference between two characterization techniques. No lattice heat capacity is subtracted from $C_\text{p}$. (E) The respective peak temperatures $T_{\text{chi}}$ (blue) and $T_\text{C}$ (red) of $\chi_{\text{DC}}(T)$ and $C_\text{p}(T)$ are plotted as a function of Fe concentration $x$. These values are consistent for ZnFe$_2$O$_4$ in the clean limit but evolve differently upon Ga doping. $T_{\text{chi}}$ drops monotonically and its kink is coincident with $\chi_{\text{DC}}(T)$ becoming symmetric around $T_{\text{chi}}$ for $x < 0.95$. $T_\text{C}$ has an oscillating behavior around $T_{\text{chi}}$. All data plotted here are measured from 1000 °C grown crystals. The inset shows the *B*-site pyrochlore sub-lattice, with the connection between Fe-Fe nearest neighbors always pointing along the (1, 1, 0) family of directions.

**Fig. 2. Evolution of the imaginary AC magnetic susceptibility.** $\chi''(T)$ were measured for pure ZnFe$_2$O$_4$ grown from (A) $T_\text{g}$=850 °C and (B) $T_\text{g}$=1000 °C respectively, and Zn(Fe$_x$Ga$_{1-}$

$_x)_2O_4$ with $x$ of (C) 0.97, (D) 0.95, (E) 0.90, (F) 0.86, (G) 0.77, and (H) 0.69. The peak feature of $\chi''(T)$ at $T_{chi}$ rises with increasing inversion disorder from (A) to (B), increases with Ga-doping in **c**, and transforms to a cliff-shaped step function in (H). The bump shaped spectral weight at 4 K likely is as a feature of the antiferromagnetic state that disappears at $x\sim0.90$. Insets provide detailed regions near $T_{chi}$ to demonstrate frequency-dependent features of spin glass.

**Fig. 3. Neutron magnetic diffuse scattering of both fluctuations and static spin correlation.** Incremental changes of spin diffuse scattering intensity in the ($H$, $K$, 0) plane are plotted as $I_{T1-T2}(\boldsymbol{q}) = I_{T1}(\boldsymbol{q}) - I_{T2}(\boldsymbol{q})$ between neighboring temperatures $T1$ and $T2$ (Methods). The red (blue) color indicates accumulation (reduction) of magnetic diffuse scattering beyond $10^{-12}$ seconds at the lower temperature. (A) In $ZnFe_2O_4$, fluctuations in the interstitial region such as (0.75, 0.75, 0) are repopulated to main magnetic Bragg peaks at low temperature, showing regions of both blue and red. (B) In $Zn(Fe_{0.97}Ga_{0.03})_2O_4$, $I_{4.8-7.5K}(\boldsymbol{q})$ also demonstrates spectral weight repopulation due to fluctuations, but at lower temperatures. (C) In $Zn(Fe_{0.86}Ga_{0.14})_2O_4$, there is only accumulations of spin correlation, leading to all red regions. $I_{T1-T2}(\boldsymbol{q})$ in the ($H$, $K$, 1) plane are plotted in Supplementary Fig. 7. Raw diffuse spectra are plotted in Supplementary Figs. 5 and 6.

**Fig. 4. Characteristics of magnetic diffuse scattering across the reciprocal space.** Magnetic diffuse scattering intensities $I(T)$ of three crystal sets are individually integrated over spherical volumes specified by the radius $r$ around the wavevector $\boldsymbol{q}$ and averaged over all equivalent reciprocal space positions (Methods). The schematics illustrate all choices of momentum space position and volume in the ($H$, $K$, 0) plane, using data of $ZnFe_2O_4$ at 4.8 K as the template. (A) The overall magnetic spectral intensity $I(T)$ is integrated around two reflections (1, 0.5, 0) and (1, 1.5, 0) to account for the effects of magnetic structure factor (Methods). (B, C) $I(T)$ of (1, 0.5, 0) reflection is integrated over spherical volumes of various sizes. In $ZnFe_2O_4$ and $Zn(Fe_{0.97}Ga_{0.03})_2O_4$, a high intensity over a small volume indicates the long-range antiferromagnetism, and in $Zn(Fe_{0.86}Ga_{0.14})_2O_4$, a spread distribution of scattering intensity indicates the short-range order. (D) $I(T)$ is also integrated at (0.75, 0.75, 0) as a comparison (Methods). In both $ZnFe_2O_4$ and $Zn(Fe_{0.97}Ga_{0.03})_2O_4$, $I(T)$ reduces below $T_C$, indicating rearrangements of the spectral weight due to critical fluctuations. In comparison, in $Zn(Fe_{0.86}Ga_{0.14})_2O_4$, $I(T)$ increases continuously at (0.75, 0.75, 0), and the intensity evolution is very similar in magnitude to that of (1, 0.5, 0).

**Fig. 5. Temperature evolution of spin correlation lengths $\xi$ along different momentum space vectors.** (A) $\xi_W(T)$ is extracted from longitudinal scans across the $W$ point or the magnetic (1, 0.5, 0) reflection. $\xi(T)$ of the lattice (1, 1, 1) reflection is plotted alongside as a reference of the instrument resolution limit for each sample assembly. For $Zn(Fe_{0.97}Ga_{0.03})_2O_4$, the spin correlation length is not resolution-limited but finite. A vertical dashed line is drawn for each sample, representing the peak temperature position $T_C$ of the heat capacity from Fig. 1; this is correlated with the inflection point of $\xi_W(T)$. (B) $\xi_{(0.75,0.75,0)}(T)$ of the short-range order along the (1, 1, 0) direction, extracted from the longitudinal scans at the Brillouin zone boundary (0.75, 0.75, 0). $\xi_{(0.75,0.75,0)}(T)$ for $Zn(Fe_{0.86}Ga_{0.14})_2O_4$ continuously increases as temperature lowers. All solid curves are guides to the eye.

**Supplementary Fig. 1. Controlling *A*-site stoichiometry and Fe concentration during the sample growth.** (A) $1/\chi_{DC}$ of several $Zn(Fe_xGa_{1-x})_2O_4$ crystals with different Fe concentrations. The Curie-Weiss fitting generates $T_{CW}$ and the Fe concentration $x$ from the linear slope. (B) $T_{CW}$ depends on the initial mixing molar ratio $r$ of ZnO vs. $Fe_2O_3$ and $Ga_2O_3$ powders in the high temperature solution. The minimum of $T_{CW}$ indicates the condition $r_{stoich}$ to grow Zn-stoichiometric crystal at the specific $Ga_2O_3$ composition $y$. At $r_{stoich}$, $T_{CW}$ of all $Zn(Fe_xGa_{1-x})_2O_4$ crystals are about -20 K. (C) Dependence of $r_{stoich}$ to $Ga_2O_3$ composition $y$, with the dashed line as an extension. A powder ratio $y = 0.50$ in solution corresponds to $x \sim 0.86$ in grown crystals. (D) The lattice constant $a$ has a smooth and monotonic dependence on measured $x$.

**Supplementary Fig. 2. Influences of off-stoichiometry on DC magnetic susceptibility $\chi_{DC}(T)$.** For a fixed growth parameter $y$, zero-field cooled $\chi_{DC}(T)$ are plotted for several crystals of $Zn(Fe,Ga)_2O_4$ grown with different molar ratio $r$ of ZnO vs. total $Ga_2O_3$ and $Fe_2O_3$. The stoichiometric condition leads to the lowest magnitude of $\chi_{DC}(T)$.

**Supplementary Fig. 3. Neutron magnetic scattering sample assemblies.** (A) A mosaic assembly of $ZnFe_2O_4$ within a volume of $10 \times 10 \times 10$ mm$^3$. (B) $Zn(Fe_{0.97}Ga_{0.03})_2O_4$, and (C) $Zn(Fe_{0.86}Ga_{0.14})_2O_4$ are each assembled within a volume of $8 \times 8 \times 8$ mm$^3$ (Methods).

**Supplementary Fig. 4. Real and Imaginary AC magnetic susceptivity of $ZnFe_2O_4$.** Three stoichiometric $ZnFe_2O_4$ specimens were grown with temperature of (A) 1250 °C, (B) 1000 °C, and (C) 850 °C (Methods). The plotted susceptibility data are measured with a frequency of 9984 Hz. Higher growth temperatures produce higher values of $\chi'$ and $\chi''$ at $T_{chi}$, with peak values of $\chi'$ differ by a factor of almost two between 1250 °C and 850 °C grown crystals. In addition, the peak ratio $\chi''/\chi'$ at around $T_{chi}$ also grows higher, reaching ~2.2% for 1250 °C grown crystals, in comparison to ~0.4% and ~0.12% for crystals grown from 1000 °C and 850 °C respectively. A $\chi''/\chi'$ ratio ~3.3% at $T_{chi}$=13K was reported for $ZnFe_2O_4$ crystals grown from 1200 °C down [24].

**Supplementary Fig. 5. Raw neutron magnetic diffuse scattering intensities.** Intensities in the ($H$, $K$, 0) plane of all three $Zn(Fe_xGa_{1-x})_2O_4$ crystal sets, colored in a log scale.

**Supplementary Fig. 6. Raw neutron magnetic diffuse scattering intensities.** Intensities in in the ($H$, $K$, 1) plane of all three $Zn(Fe_xGa_{1-x})_2O_4$ crystal sets, colored in a log scale.

**Supplementary Fig. 7. Differential neutron magnetic diffuse scattering intensities in the ($H$, $K$, 1) plane.** These $I_{T1-T2}(q)$ plots are equivalents to Fig. 3 at $L$=1, for all three $Zn(Fe_xGa_{1-x})_2O_4$ crystal sets.

**Acknowledgement:** Y.F. is grateful for the visiting appointment at Caltech hosted by T. F. Rosenbaum, where seminal discussions with Patrick A. Lee, Thomas F. Rosenbaum, and Daniel M. Silevitch took place. We also thank Takeshi Egami for insightful discussions and Haidong Zhou for the loan of $Yb_2Sn_2O_7$ specimen during the ACMS calibration. Y.F. acknowledge financial support from the Okinawa Institute of Science and Technology Graduate University, with subsidy funding from the Cabinet Office, Government of Japan. Y.W. acknowledges support from the Department of Energy, Office of Science, Office of Basic Energy Sciences, under Award Number DE-SC0024941. ACMS measurements were partially supported by the Electromagnetic Properties Lab core facility at the University of Tennessee Knoxville. A portion of this research used resources at the Spallation Neutron Source, a U.S. Department of Energy Office of Science User Facility operated by the Oak Ridge National Laboratory. The neutron beam time was allocated to CORELLI on proposal #IPTS-32460.

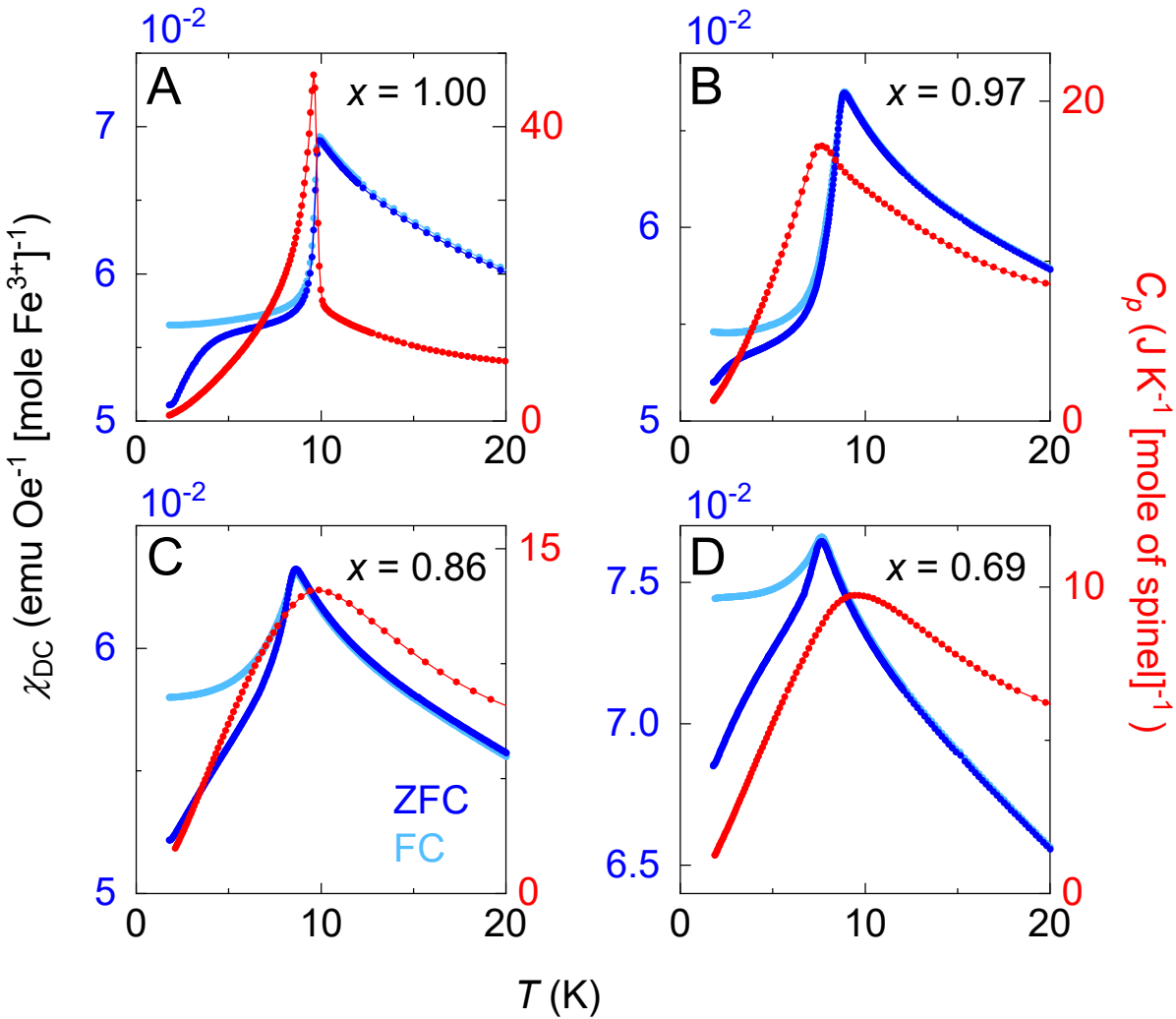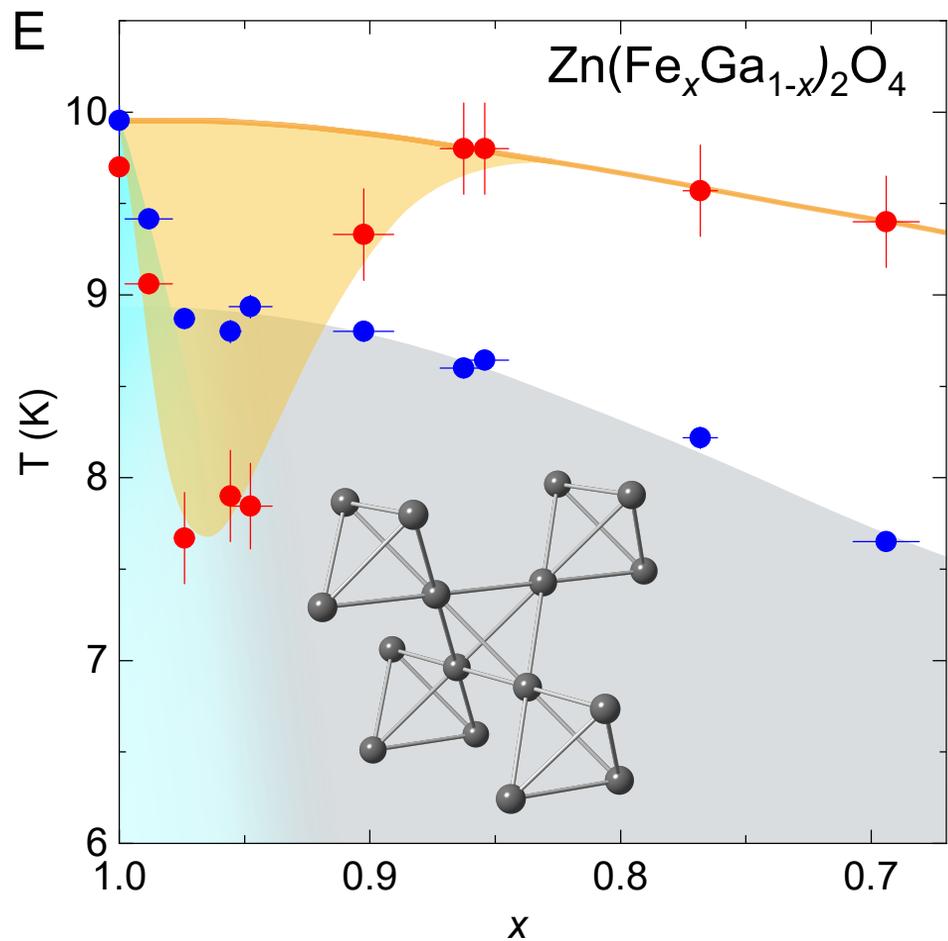

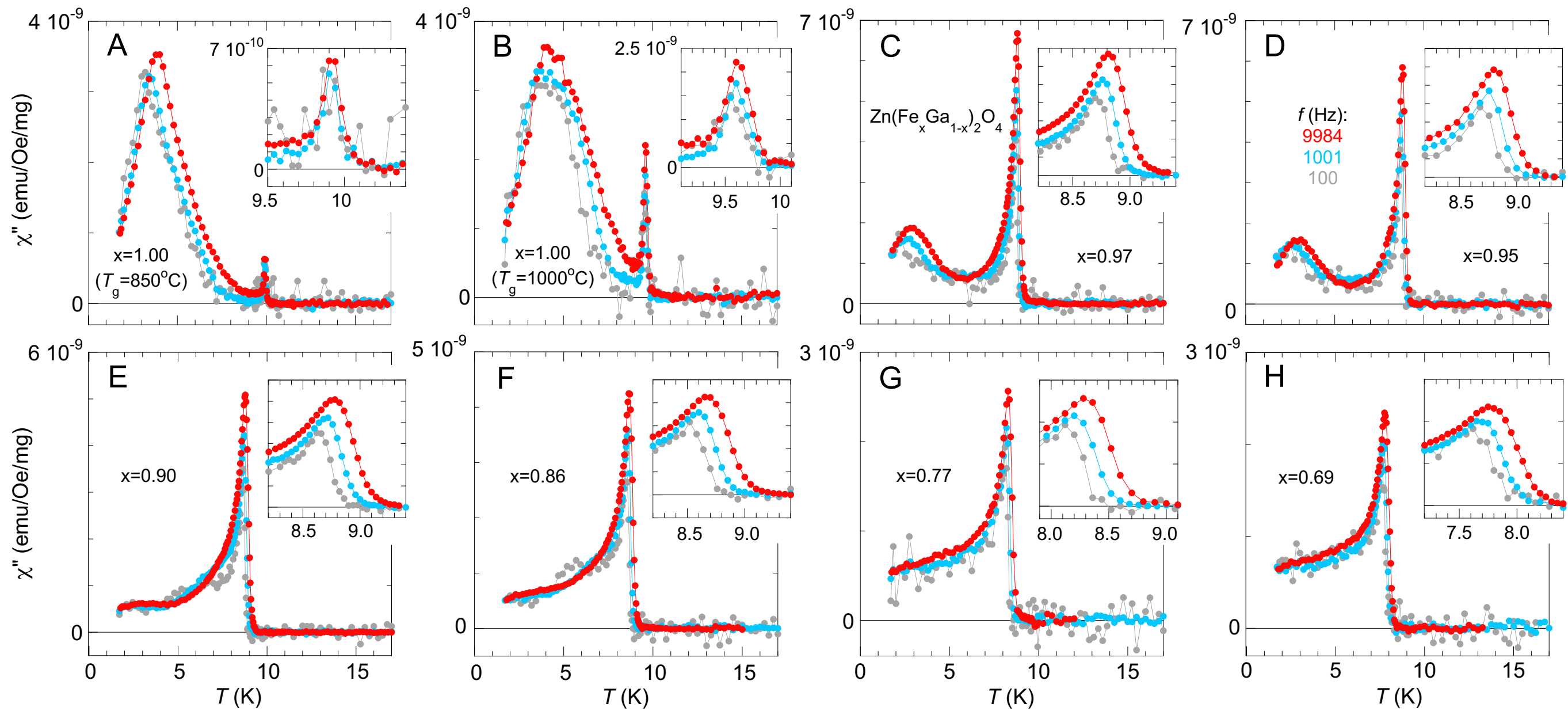

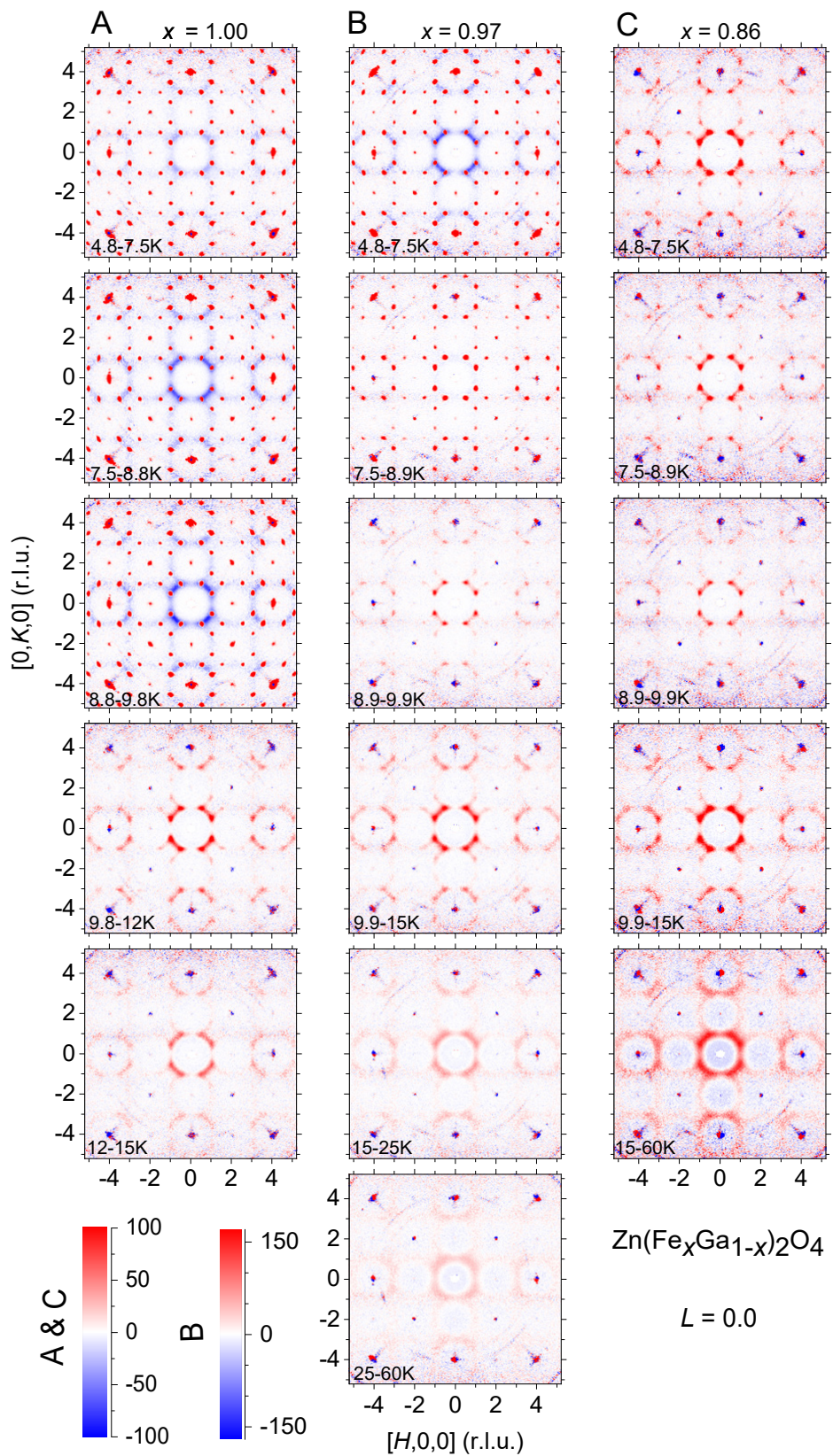

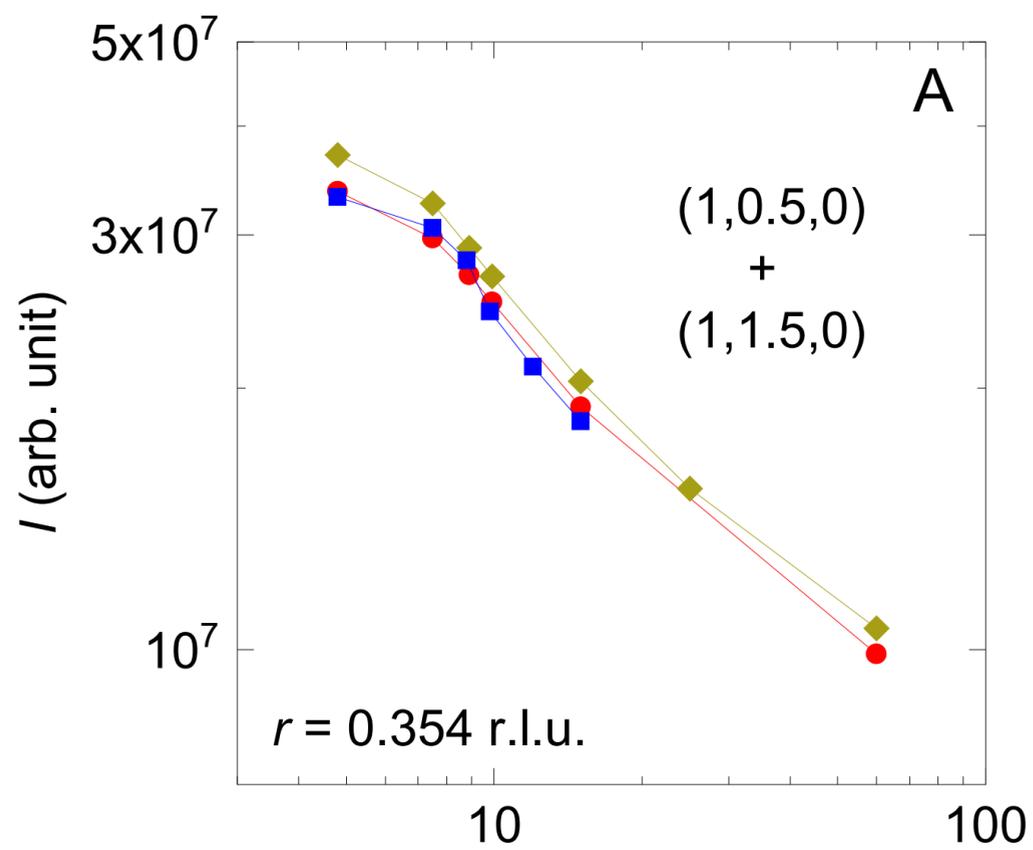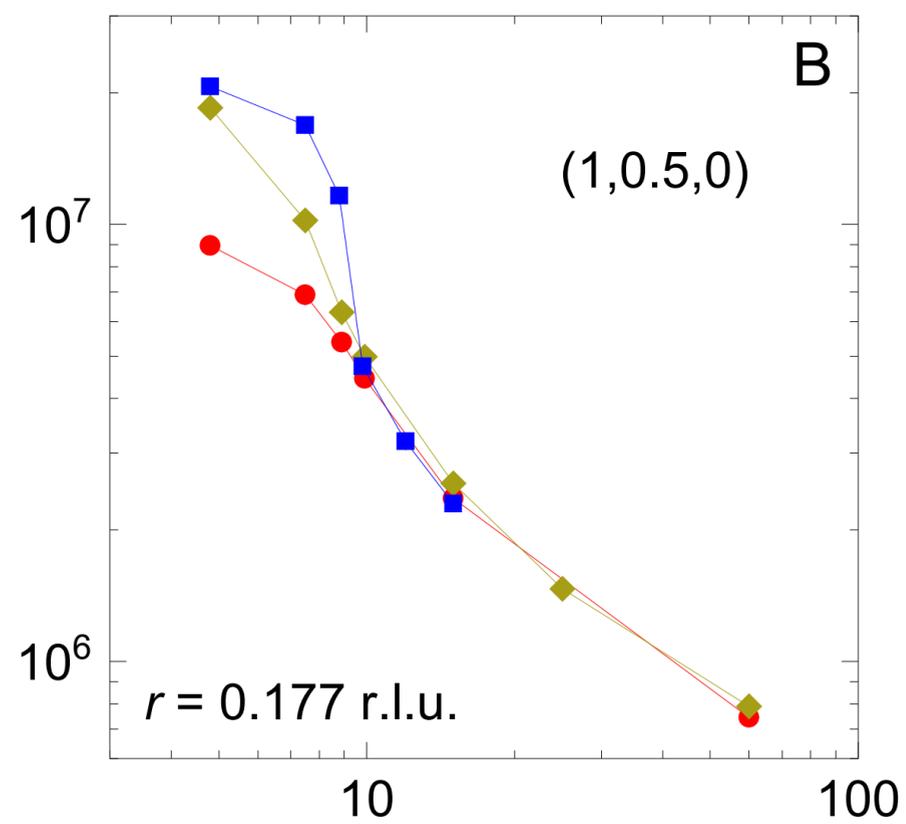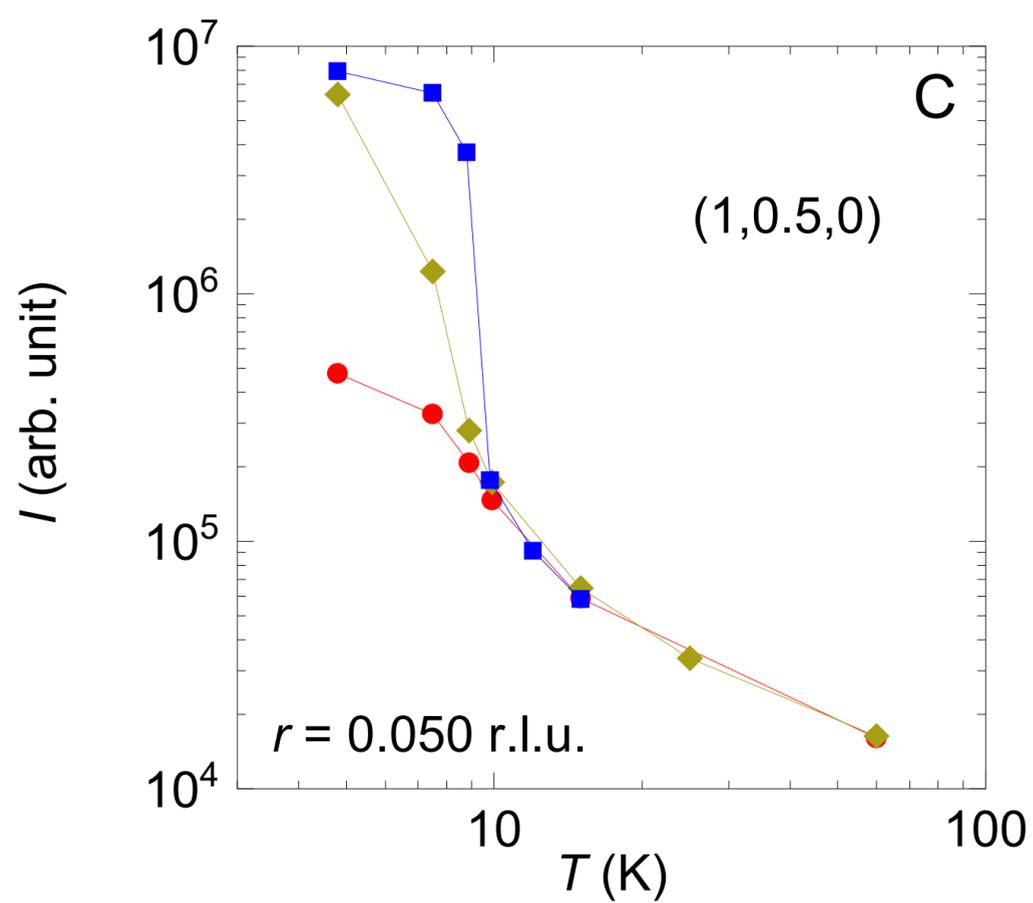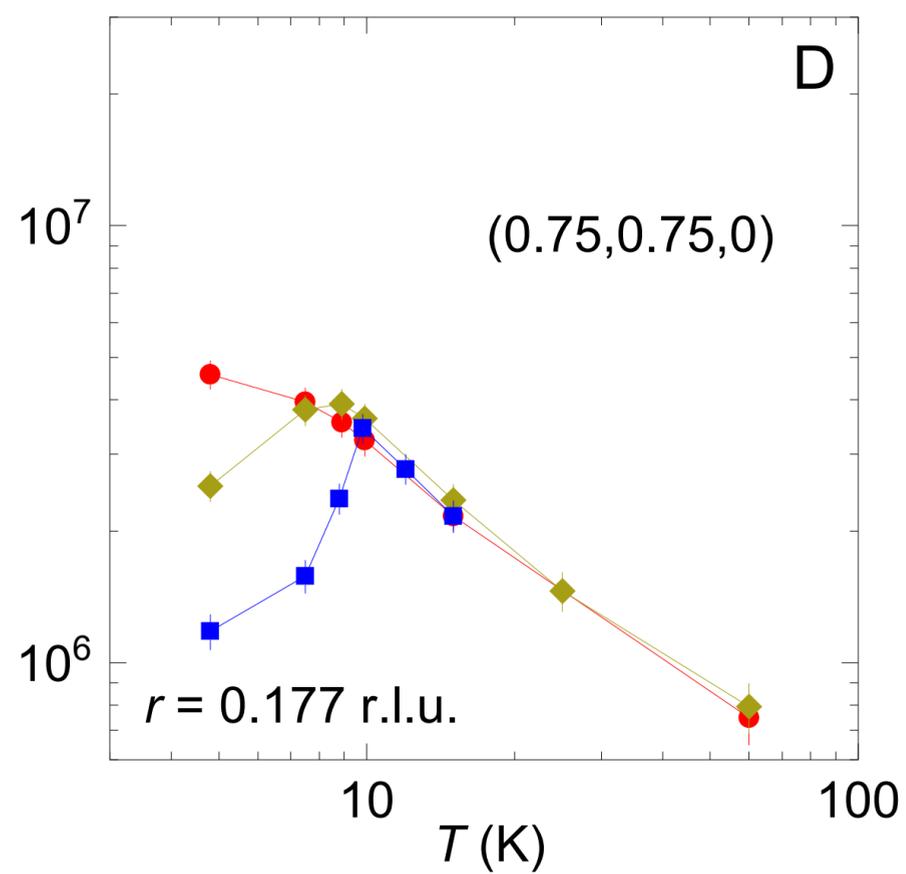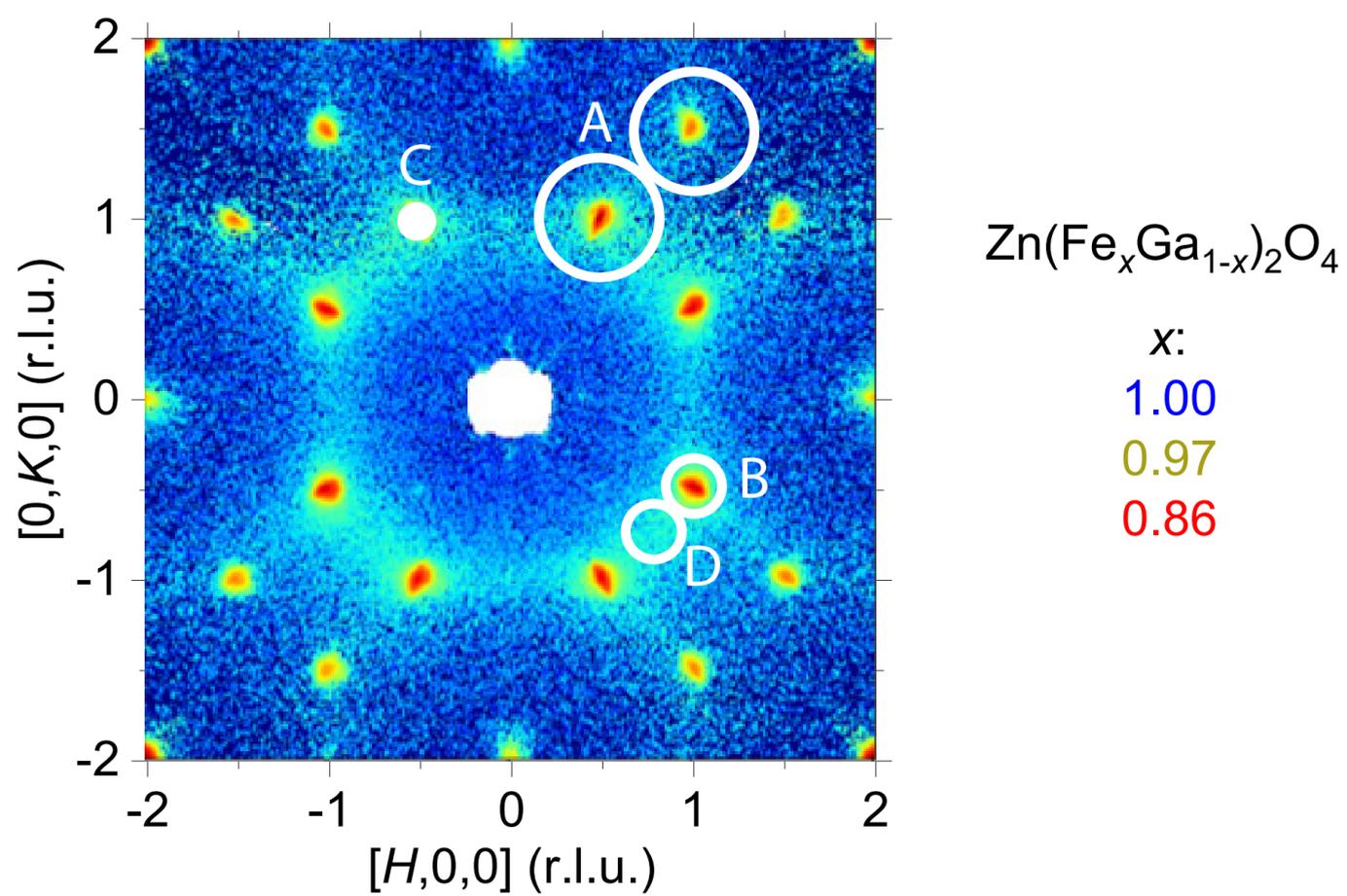

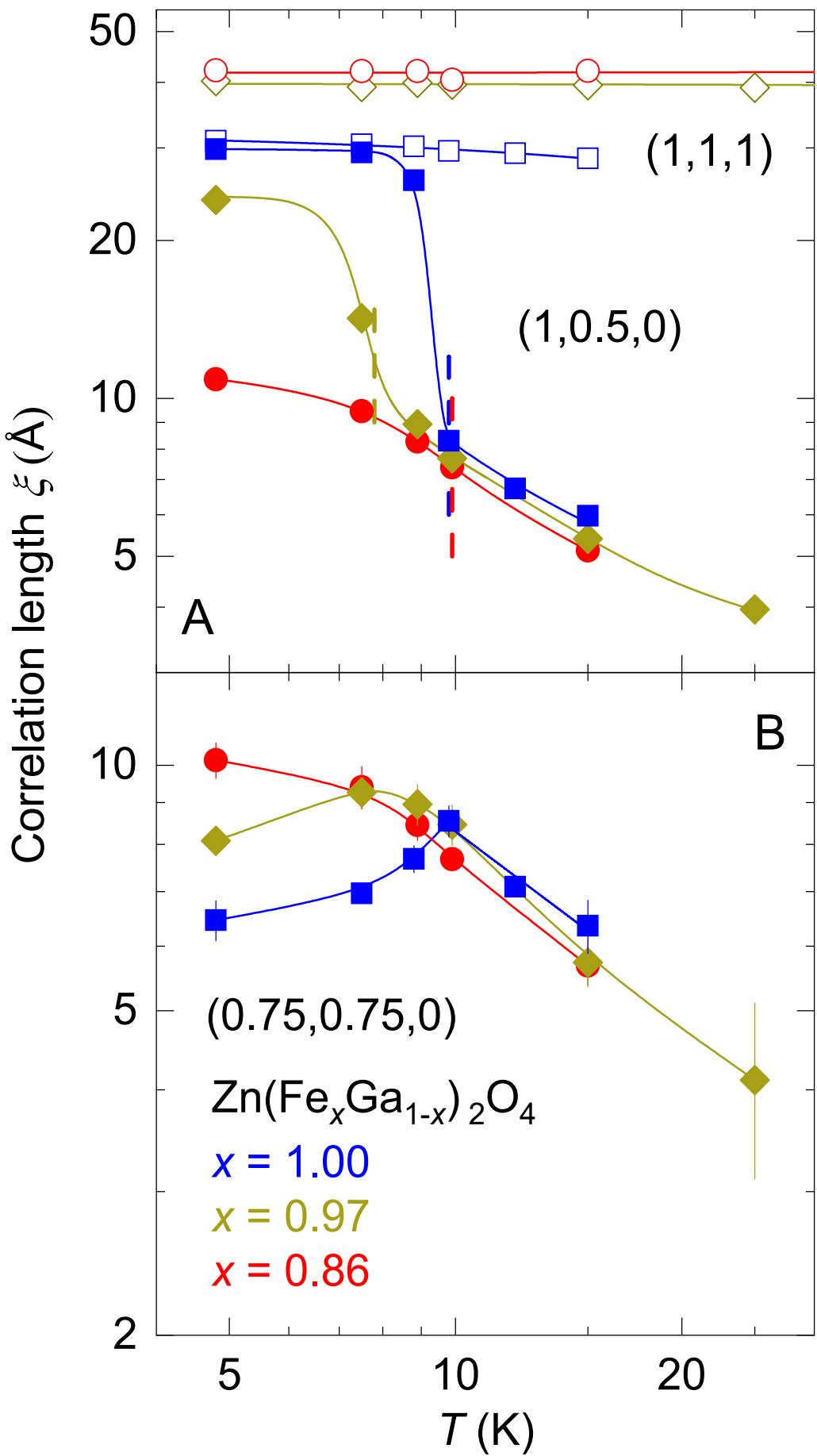

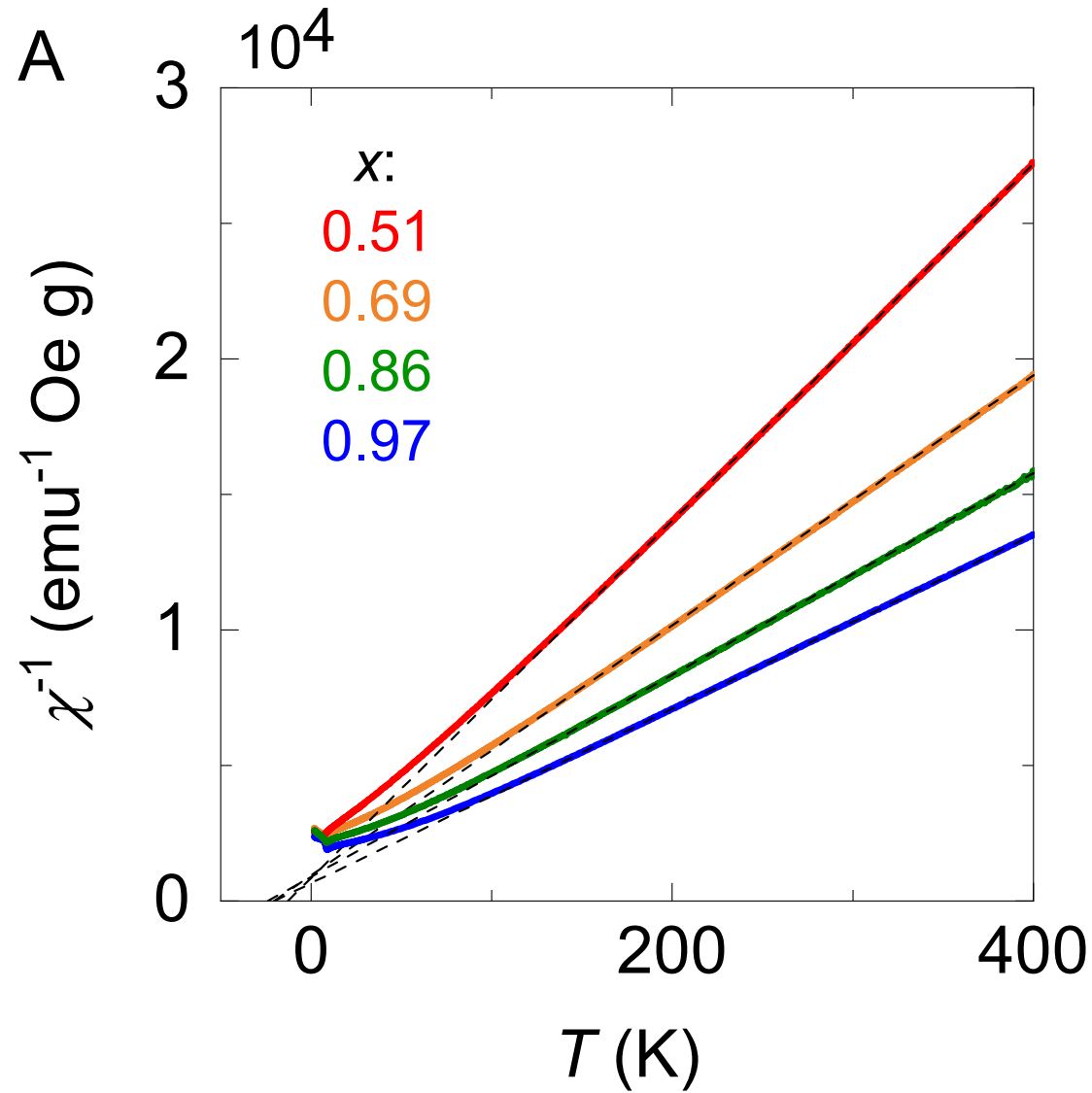
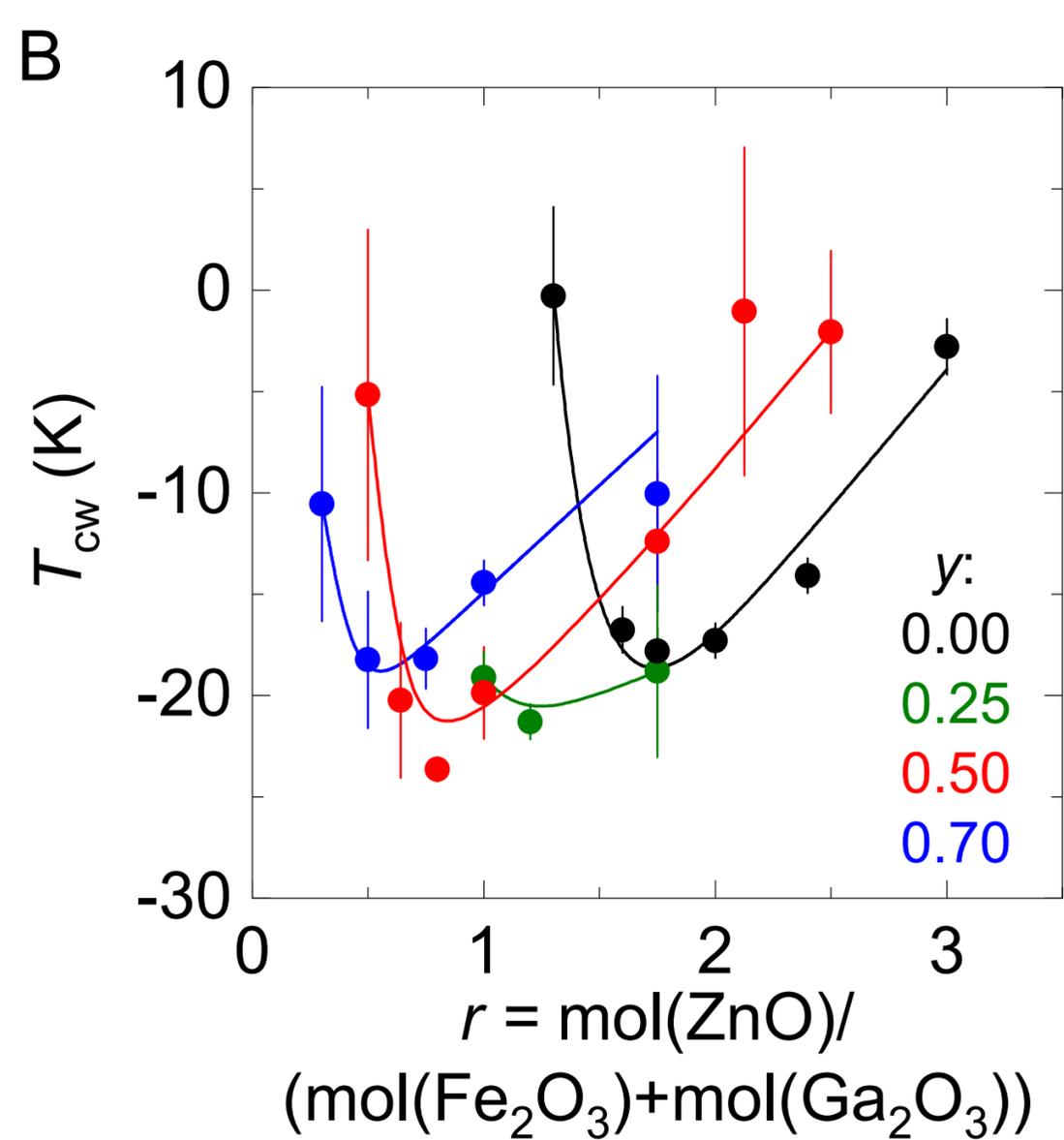
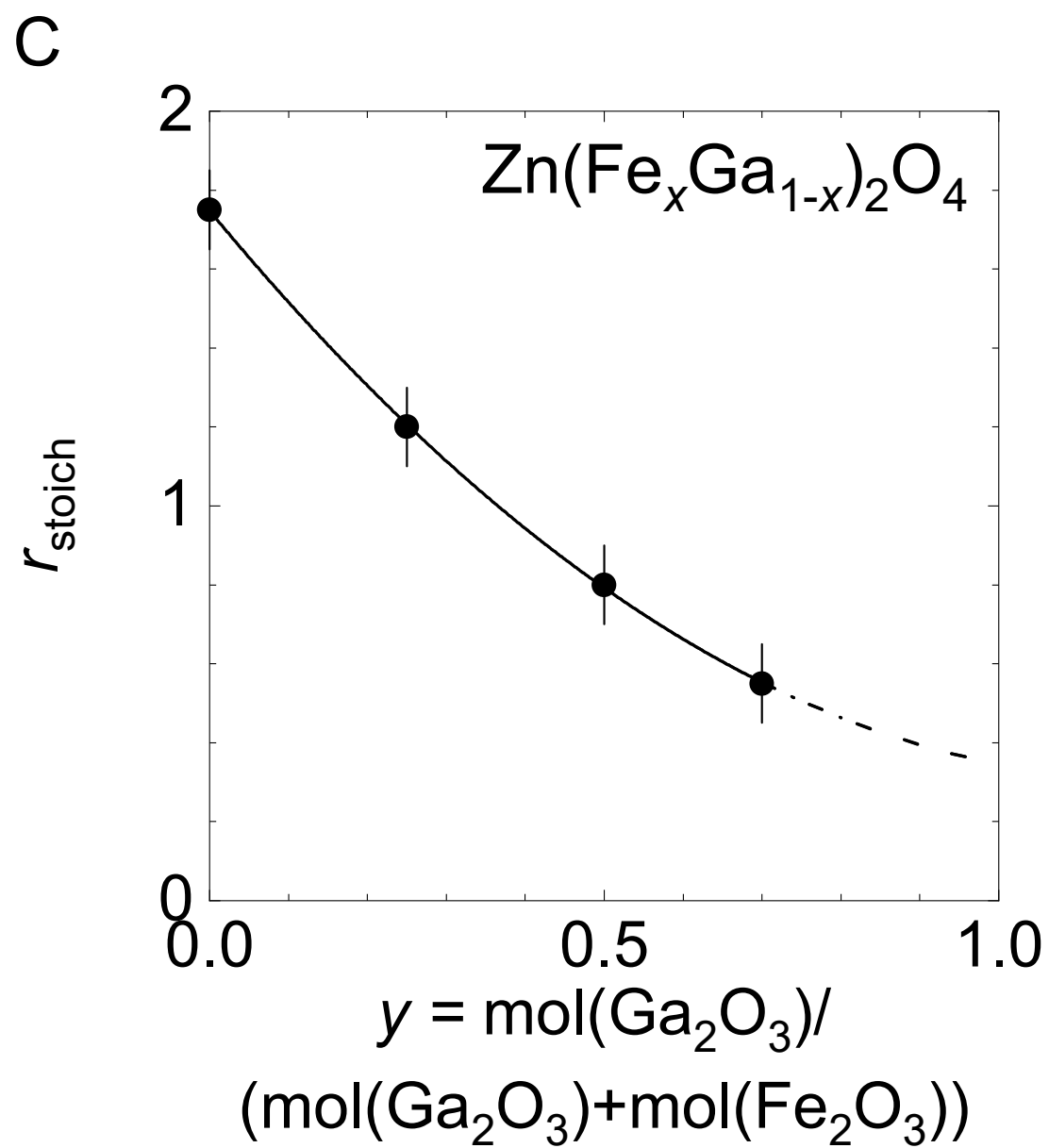
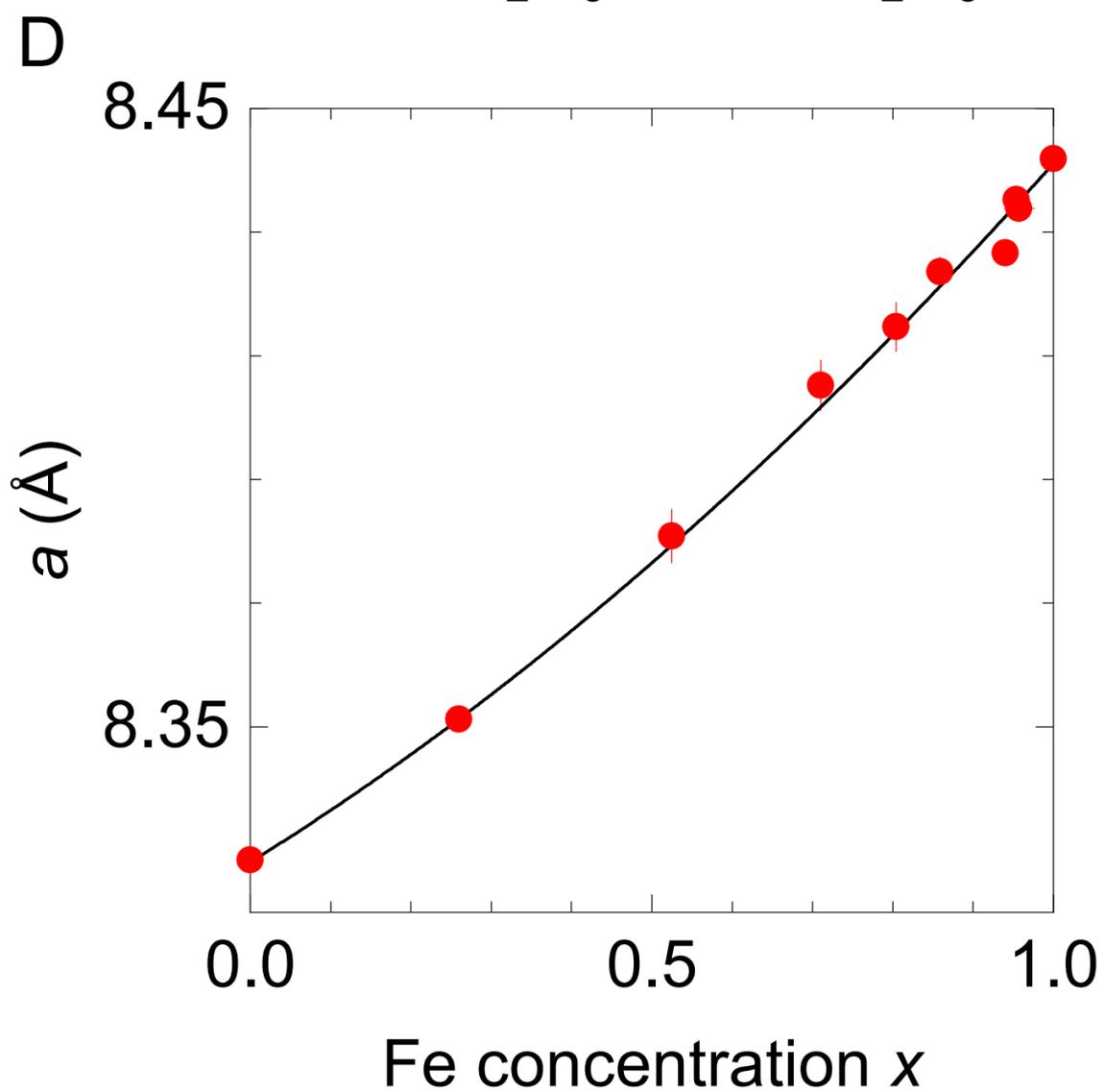

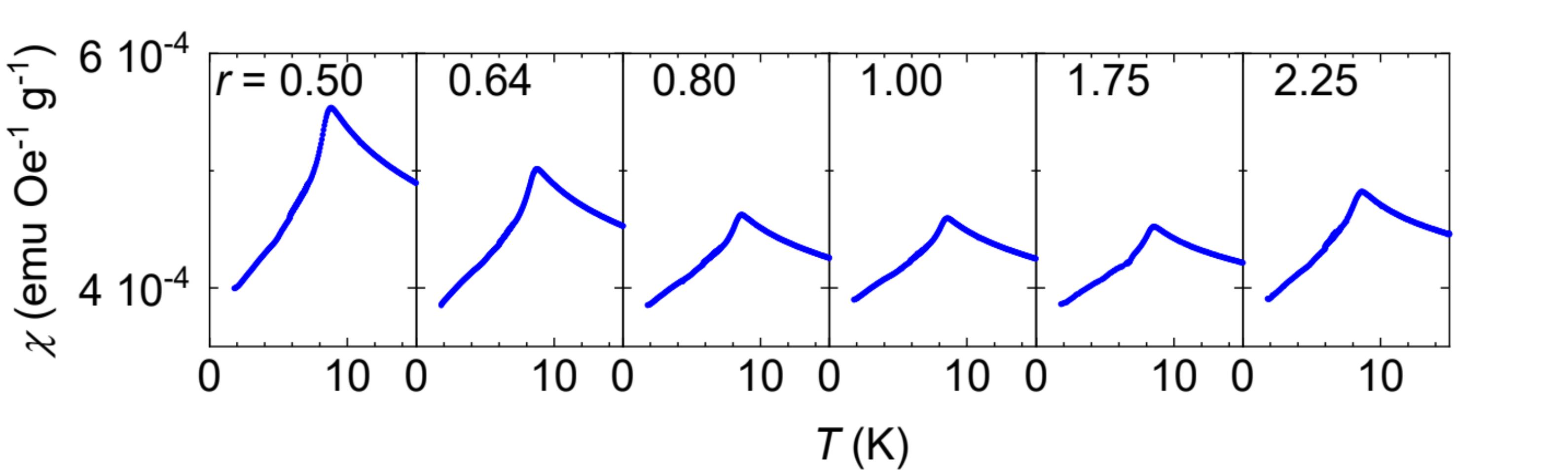

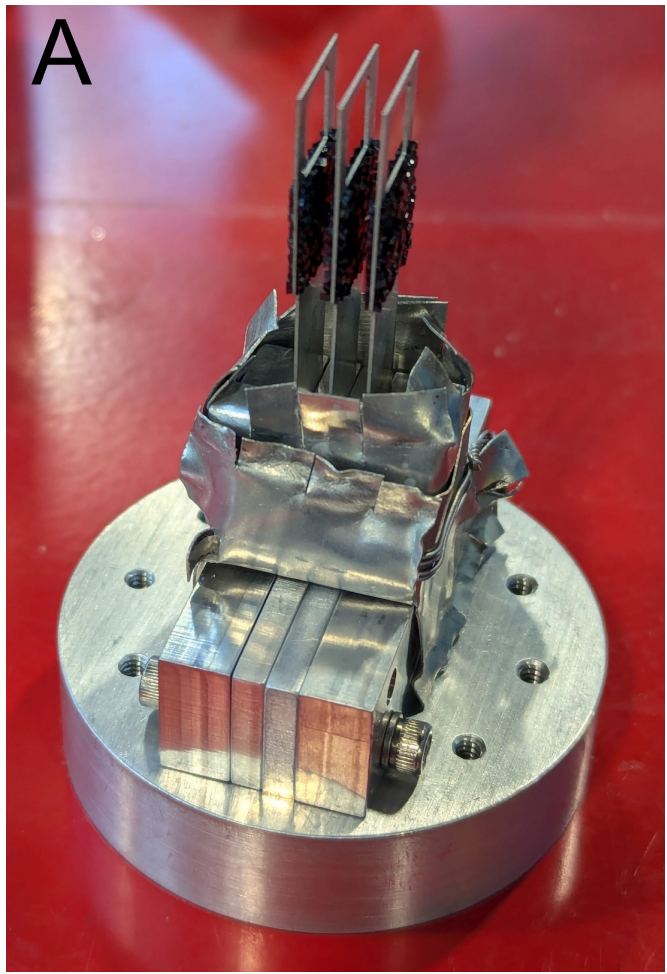 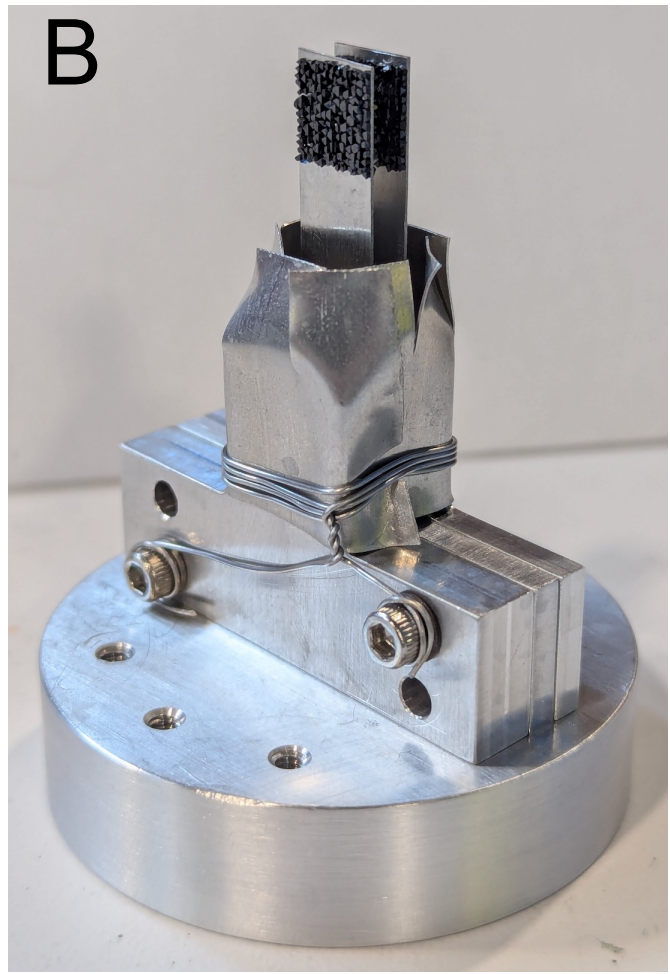 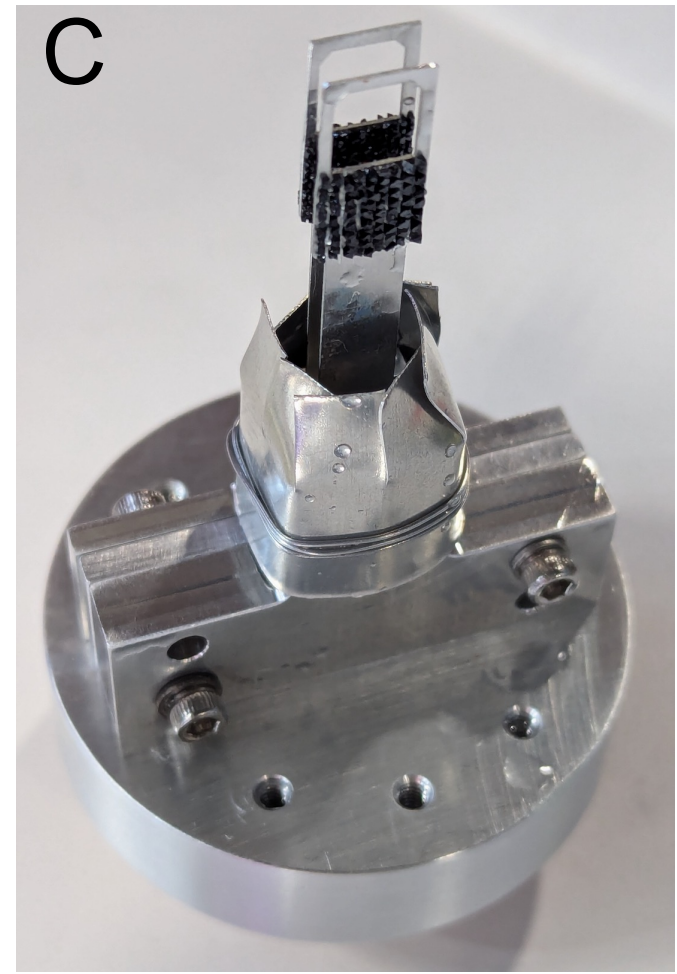

ZnFe$_2$O$_4$  |  Zn(Fe$_{0.97}$Ga$_{0.03}$)$_2$O$_4$  |  Zn(Fe$_{0.86}$Ga$_{0.14}$)$_2$O$_4$

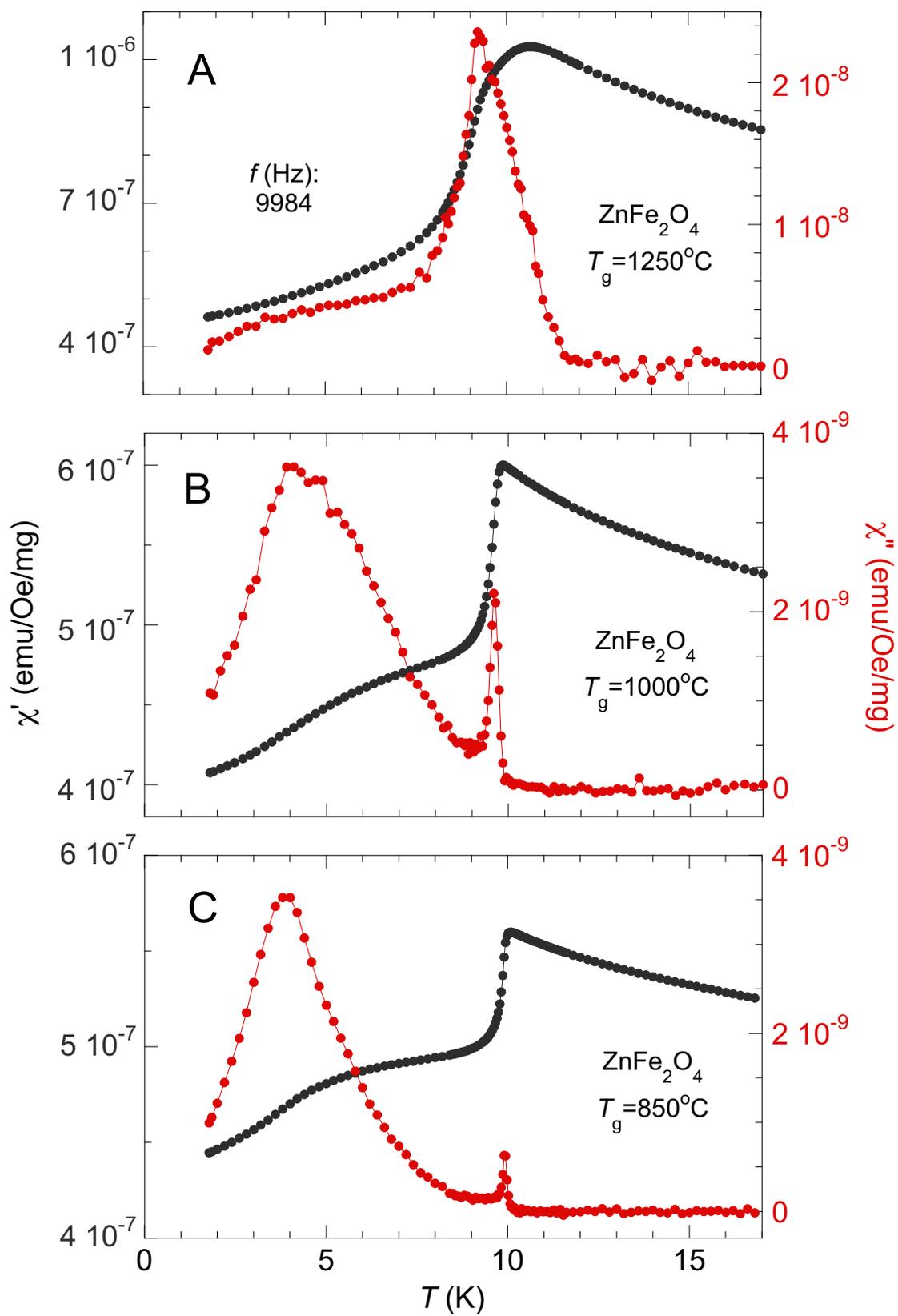

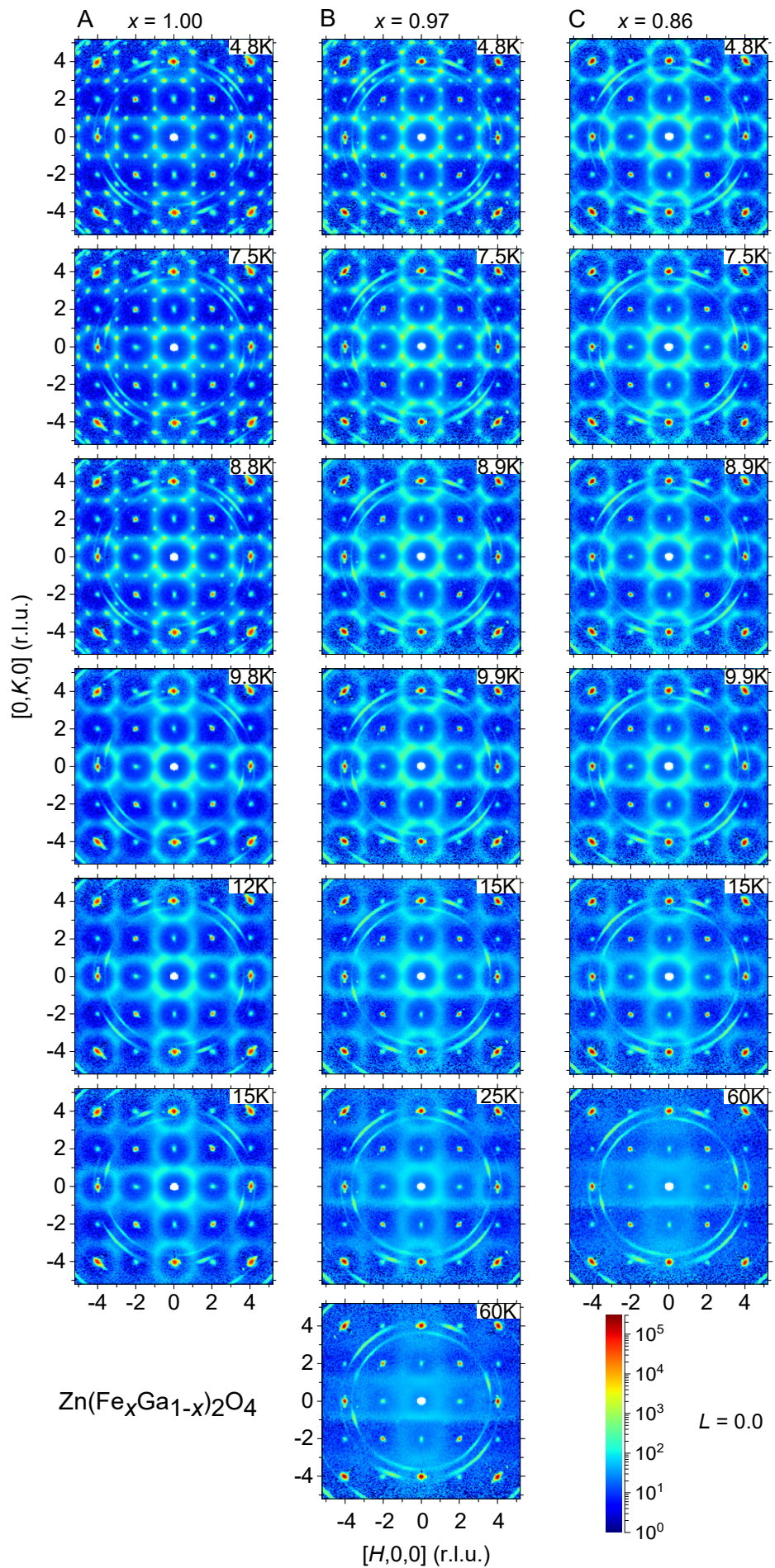

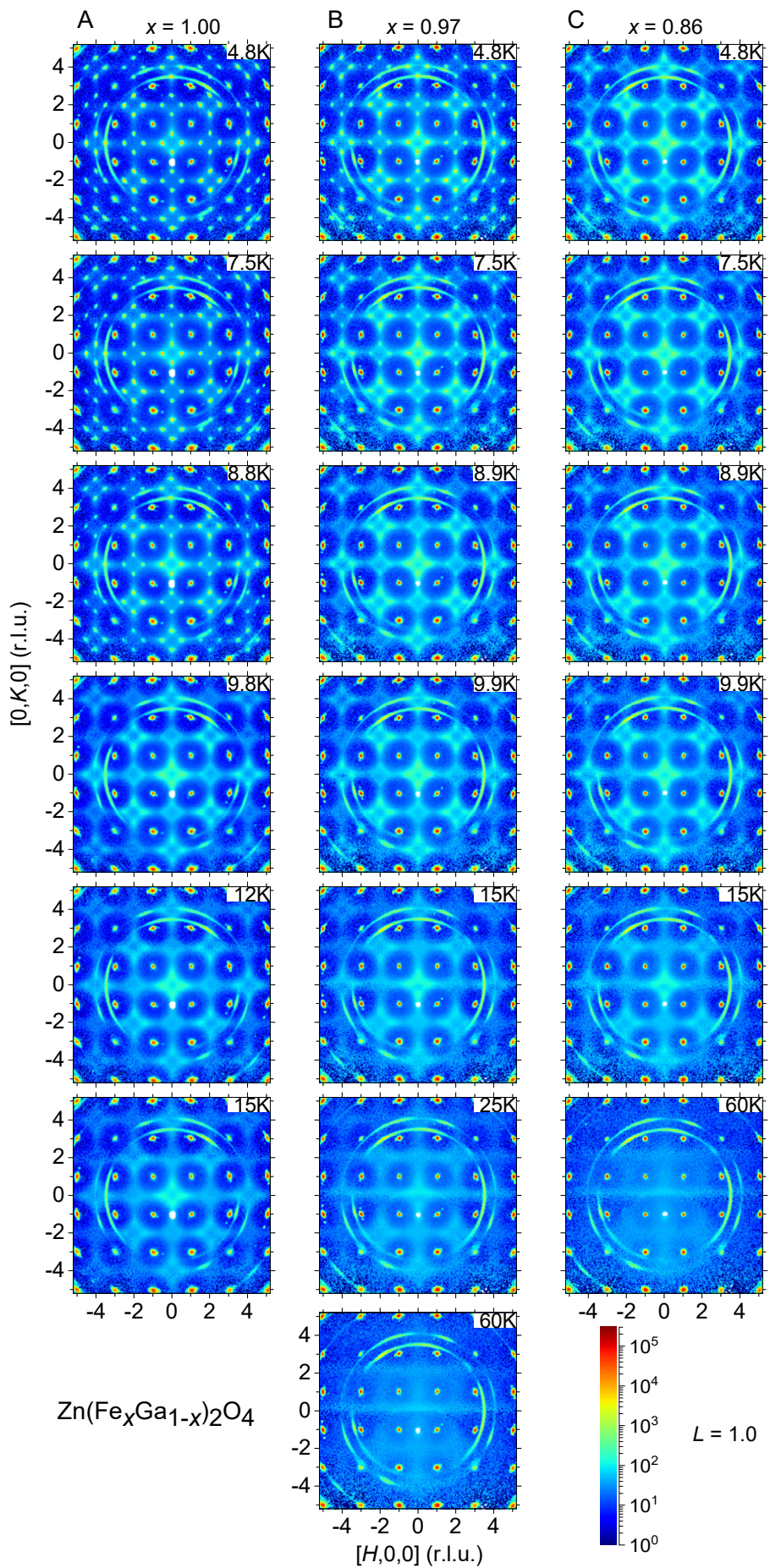

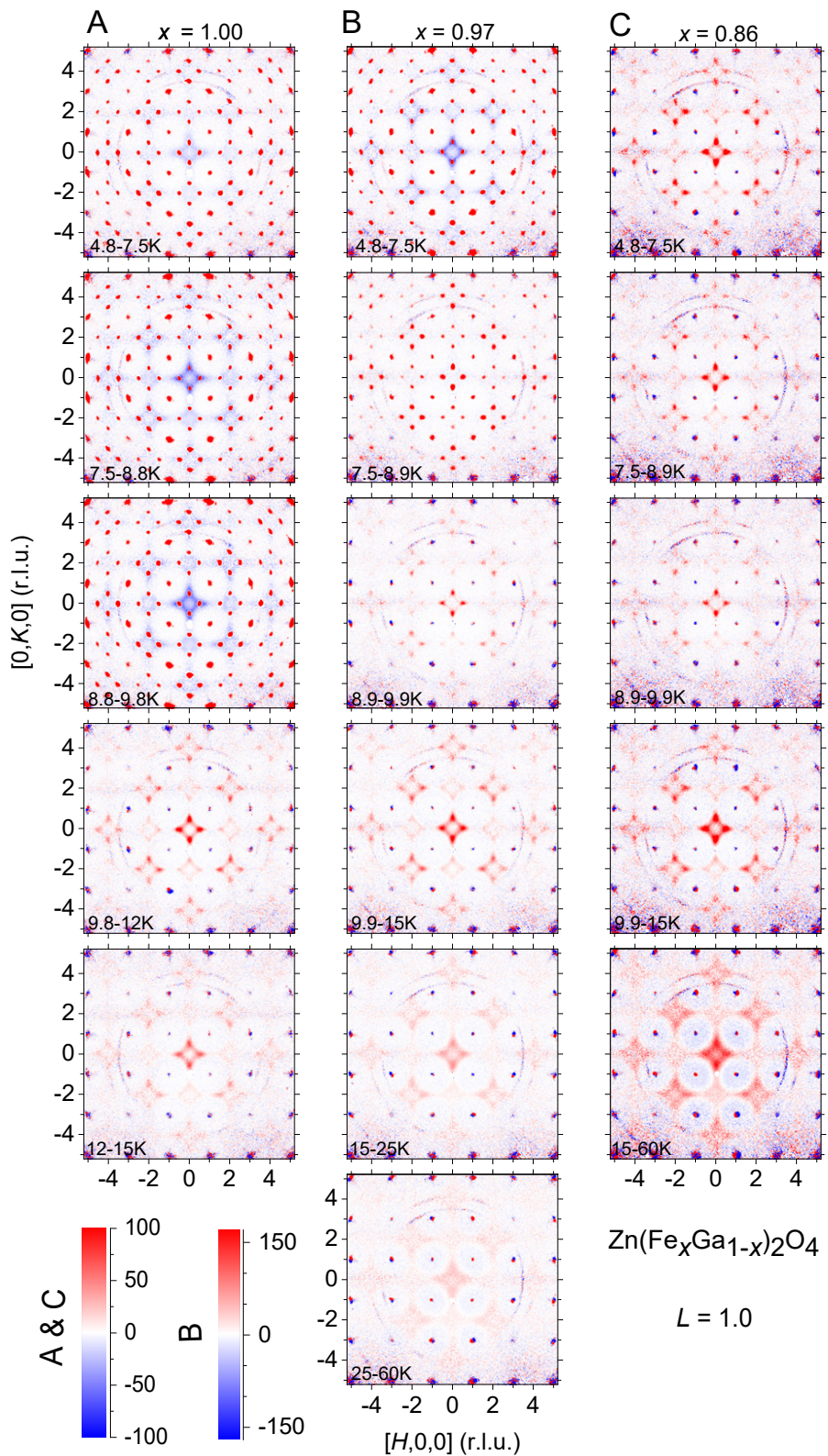